\documentstyle[aps,eqsecnum,preprint,floats,epsf,psfig]{revtex}
\begin{document}
\def\be{\begin{eqnarray}}
\def\en{\end{eqnarray}}
\def\non{\nonumber}
\def\la{\langle}
\def\ra{\rangle}
\def\ep{\varepsilon}
\def\vma{{_{V-A}}}
\def\vpa{{_{V+A}}}
\newcommand{\eps}{\varepsilon}
\def\BD{{{m_B+m_D\over 2\sqrt{m_B m_D}}}}
\def\BDstar{{{m_B+m_{D^*}\over 2\sqrt{m_B m_{D^*}}}}}
\def\BP{{{m_B+m_P\over 2\sqrt{m_B m_P}}}}
\def\BV{{{m_B+m_V\over 2\sqrt{m_B m_V}}}}
\def\ccV{{\sqrt{\frac{2}{\omega_{BV}(q^2)+1}}}}
\def\ccVin{{\sqrt{\frac{\omega_{BV}(q^2)+1}{2}}}}
\def\ccP{{\sqrt{\frac{2}{\omega_{BP}(q^2)+1}}}}
\def\ccPin{{\sqrt{\frac{\omega_{BP}(q^2)+1}{2}}}}
\def\K{{K^{(*)}}}
\def\D{{D^{(*)}}}
\newcommand{\B}{{\mathcal{B}}}
\newcommand{\F}{{\mathcal{F}}}
\newcommand{\G}{{\mathcal{G}}}
\def\up{\uparrow}
\def\dw{\downarrow}
\def\non{\nonumber}
\def\la{\langle}
\def\ra{\rangle}
\def\ov{\overline}
\def\nc{N_c^{\rm eff}}
\def\vp{\varepsilon}
\def\vcbud{{V_{cb}V_{ud}^*}}
\def\vcbcs{{V_{cb}V_{cs}^*}}
\def\C1{{\left( \frac{G_F}{\sqrt{2}}\vcbud\right)^2}}
\def\r{{\tau(B^-)\over\tau(\overline{B}^0)}}
\def\pr{{\sl Phys. Rev.}~}
\def\prl{{\sl Phys. Rev. Lett.}~}
\def\pl{{\sl Phys. Lett.}~}
\def\np{{\sl Nucl. Phys.}~}
\def\zp{{\sl Z. Phys.}~}

\draft
\title{\large\bf Updated Analysis of $a_1$ and $a_2$\\
 in Hadronic Two-body Decays of $B$ Mesons}
\vskip 1cm
\author{Hai-Yang Cheng \footnote{Email address:
{\tt phcheng@ccvax.sinica.edu.tw}} and Kwei-Chou Yang
\footnote{Email address: {\tt kcyang@phys.sinica.edu.tw}}}
\address{Institute of Physics, Academia Sinica, Taipei, Taiwan 115,
Republic of China}
\maketitle
\date{June 1998}
%%%%%%%%%%%%%%%%%%%%%%%%%%%%%%%%%%%%%%%%%%%%%%%%%%%%%%%%%%%%%%%%%%%%%%%%%%
\begin{abstract}
Using the recent experimental data of $B\to D^{(*)}
(\pi,\rho)$, $B\to D^{(*)} D_s^{(*)}$, $B\to J/\psi K^{(*)}$
and various model calculations on form factors, we re-analyze
the effective coefficients $a_1$ and $a_2$ and their ratio. QCD and
electroweak penguin corrections to $a_1$ from $B\to D^{(*)}D_s^{(*)}$ and
$a_2$ from $B\to J/\psi K^{(*)}$ are estimated.
In addition to the model-dependent determination,
the effective coefficient $a_1$ is also extracted in a model-independent way
as the decay modes $B\to D^{(*)}h$ are related by factorization
to the measured semileptonic distribution of $B\to D^{(*)}\ell\bar\nu$ at
$q^2=m_h^2$. Moreover, this enables us to extract model-independent
heavy-to-heavy form factors, for example, $F_0^{BD}(m_\pi^2)=0.66\pm 0.06\pm
0.05$ and $A_0^{BD^*}(m_\pi^2)=0.56\pm 0.03\pm 0.04$.
The determination of the magnitude of $a_2$ from $B\to J/\psi
K^{(*)}$ depends on the
form factors $F_1^{BK}$, $A_{1,2}^{BK^*}$ and $V^{BK^*}$ at $q^2=m^2_{J/
\psi}$. By requiring that $a_2$ be process insensitive ( i.e., the value of
$a_2$ extracted from $J/\psi K$ and $J/\psi K^*$ states should be similar), as
implied by the factorization hypothesis,
we find that $B\to K^{(*)}$ form factors are severely constrained; they
respect the relation
$F_1^{BK}(m^2_{J/\psi})\approx 1.9 A_1^{BK^*}(m^2_{J/\psi})$.
Form factors $A_2^{BK^*}$ and $V^{BK^*}$ at $q^2=m^2_{J/\psi}$ inferred
from the measurements of the longitudinal polarization fraction and the
$P$--wave component in $B\to J/\psi K^*$ are obtained.
A stringent upper limit on $a_2$ is derived from the current bound on
$\ov B^0\to D^0\pi^0$ and it is sensitive to final-state interactions.

\end{abstract}
%\pagebreak
\section{Introduction}
Nonleptonic two-body decays of $B$ and $D$ mesons have been
conventionally studied in the generalized factorization approach
in which the decay amplitudes are approximated by the factorized hadronic
matrix elements multiplied by some universal, process-independent effective
coefficients $a_i^{\rm eff}$. Based on the generalized
factorization assumption, one can catalog
the decay processes into three classes. For class-I decays, the
decay amplitudes, dominated by the color-allowed external $W$-emission, are
proportional to $a_1^{\rm eff}\la O_1\ra_{\rm fact}$ where $O_1$ is a charged
current--charged current 4-quark operator. For class-II decays, the
decay amplitudes, governed by the color-suppressed internal
$W$-emission, are described by $a_2^{\rm eff}\la O_2\ra_{\rm fact}$ with $O_2$
being a neutral current--neutral current 4-quark operator. The decay
amplitudes of the class-III decays involve a linear combination of
$a_1^{\rm eff}
\la O_1\ra_{\rm fact}$ and $a_2^{\rm eff}\la O_2\ra_{\rm fact}$. If
factorization works, the effective coefficients
$a_i^{\rm eff}$ in nonleptonic $B$ or $D$ decays should be channel by channel
independent. Since the factorized hadronic matrix elements $\la O_i
\ra_{\rm fact}$ are renormalization scheme and scale independent, so are
$a_i^{\rm eff}$.

    What is the relation between the effective coefficients $a_i^{\rm eff}$
and the Wilson
coefficients in the effective Hamiltonian approach ? Under the naive
factorization hypothesis, one has
\be
a_1(\mu)=c_1(\mu)+{1\over N_c}c_2(\mu), \qquad \quad a_2(\mu)=c_2(\mu)+{1\over
N_c}c_1(\mu),
\en
for decay amplitudes induced by current-current operators $O_{1,2}(\mu)$,
where $c_{1,2}(\mu)$ are the corresponding Wilson coefficients.
However, this naive factorization approach encounters two principal
difficulties: (i) the
above coefficients $a_i$ are scale dependent, and (ii) it fails to describe
the color-suppressed class-II decay modes. For example, the predicted
decay rate of $D^0\to \ov K^0\pi^0$ by naive factorization is too small
compared to experiment. Two different approaches have been advocated in the
past for solving the aforementioned scale problem associated with the
naive factorization approximation. In the first approach, one incorporates
nonfactorizable effects into the effective coefficients
\cite{Cheng94,Cheng96,Soares}:
\begin{eqnarray}  \label{a12}
a_1^{\rm eff} = c_1(\mu) + c_2(\mu) \left({1\over N_c}
+\chi_1(\mu)\right)\,, \qquad \quad
a_2^{\rm eff} = c_2(\mu) + c_1(\mu)\left({1\over N_c} + \chi_2(\mu)\right)\,,
\end{eqnarray}
where nonfactorizable terms are characterized by the parameters $\chi_i$.
Considering the decay $\ov B^0 \to D^+ \pi^-$ as an
example, $\chi_1$ is given by
\begin{eqnarray} \label{chi}
\chi_1(\mu)= \eps_8^{(BD,\pi)}(\mu)+{a_1(\mu)\over c_2(\mu)}\eps_1^{(BD,\pi)}
(\mu),
\end{eqnarray}
where
\begin{eqnarray}   \label{epsilon}
\eps_1^{(BD,\pi)}&&=\frac{\langle D^+ \pi^-| (\bar cb)_\vma(\bar du)_\vma| \ov
B^0\rangle} {\langle D^+| (\bar cb)_\vma| \ov B^0 \rangle \langle
\pi^-|(\bar du)_\vma|0\rangle} -1\nonumber\,, \\
\eps_8^{(BD,\pi)} &&=\frac{\langle
D^+ \pi^-| {1\over 2} (\bar c \lambda^a b)_\vma(\bar d\lambda^a u)_\vma|
\ov B^0 \rangle} {\langle D^+| (\bar cb)_\vma | \ov B^0 \rangle
\langle \pi^-|(\bar du)_\vma|0\rangle}\,,
\end{eqnarray}
are nonfactorizable terms originated from color signlet-singlet and
octet-octet currents, respectively, $(\bar q_1q_2)_\vma\equiv \bar q_1
\gamma_\mu(1-\gamma_5)q_2$, and $(\bar q_1\lambda^a q_2)_\vma
\equiv \bar q_1\lambda^a \gamma_\mu(1-\gamma_5)q_2$.
The $\mu$ dependence of the Wilson coefficients is assumed to be exactly
compensated by that of $\chi_i(\mu)$ \cite{ns}. That is, the correct $\mu$
dependence of the matrix elements is restored by $\chi_i(\mu)$.
In the second approach, it is
postulated that the hadronic matrix element $\la O(\mu)\ra$ is related to
the tree-level one via the relation $\la O(\mu)\ra=g(\mu)\la O\ra_{\rm tree}$
and that $g(\mu)$ is independent of the external hadron states. Explicitly,
\be \label{tree}
c(\mu)\la O(\mu)\ra=c(\mu)g(\mu)\la O\ra_{\rm tree}\equiv c^{\rm eff}
\la O\ra_{\rm tree}.
\en
Since the tree-level matrix element $\la O\ra_{\rm tree}$ is renormalization
scheme and scale independent, so are the effective Wilson coefficients
$c_i^{\rm eff}$ and the effective parameters $a_i^{\rm eff}$ expressed by
\cite{Ali,CT98}
\begin{eqnarray} \label{aeff}
a_1^{\rm eff} = c_1^{\rm eff} + c_2^{\rm eff} \left({1\over N_c}
+\chi_1\right)\,, \qquad \quad
a_2^{\rm eff} = c_2^{\rm eff} + c_1^{\rm eff}\left({1\over N_c} +
\chi_2\right)\,.
\end{eqnarray}

   Although naive factorization does not work in general, we still have a
new factorization scheme in which the decay amplitude is expressed in terms
of factorized hadronic matrix elements multiplied by the universal effective
parameters $a_{1,2}^{\rm eff}$ provided that $\chi_{1,2}$ are universal
(i.e. process independent) in
charm or bottom decays. Contrary to the
naive one, the improved factorization scheme does incorporate nonfactorizable
effects in a process independent form. For example, $\chi_1=\chi_2=-{1
\over 3}$ in the large-$N_c$ approximation of factorization.
Theoretically, it is clear from Eqs. (\ref{chi}) and (\ref{epsilon})
that {\it a priori} the nonfactorized terms
$\chi_i$ are not necessarily channel independent. In fact, phenomenological
analyses of two-body decay data of $D$ and $B$ mesons indicate that while
the generalized factorization hypothesis in general works reasonably well,
the effective parameters $a_{1,2}^{\rm eff}$ do show some variation from
channel to channel, especially for the weak decays of charmed mesons
 \cite{Cheng94,Kamal96}.
However, in the energetic two-body $B$ decays, $\chi_i$ are
expected to be process insensitive as supported by data \cite{ns}.

  The purpose of the present paper is to provide an updated analysis of the
effective coefficients $a_1^{\rm eff}$ and $a_2^{\rm eff}$ from
various Cabibbo-allowed two-body decays of $B$ mesons: $B\to D^{(*)}D_s^{(*)},
\,D^{(*)}(\pi,\rho),\,J/\psi K^{(*)}$. It is known that the parameter
$|a_1^{\rm eff}|$ can be extracted from $\ov B^0\to D^{(*)+}(\pi^-,\rho^-)$
and $B_s\to D^{(*)}D_s^{(*)}$, $|a_2^{\rm eff}|$ from $B\to J/\psi K^{(*)}$,
$\ov B^0\to D^{(*)0}\pi^0(\rho^0)$, and $a^{\rm eff}_2/a^{\rm eff}_1$ from
$B^-\to D^{(*)}(\pi,\rho)$. However, the determination of $a_1^{\rm eff}$ and
$a_2^{\rm eff}$ is subject to many uncertainties:
decay constants, form factors
and their $q^2$ dependence, and the quark-mixing matrix element $V_{cb}$.
It is thus desirable to have an objective estimation of $a^{\rm eff}_{1,2}$.
A model-independent extraction of $a_1$ is possible because the decay
modes $B\to D^{(*)}h$ can be related by factorization
to the measured semileptonic decays $B\to D^{(*)}\ell\bar\nu$. As a
consequence, the ratio of nonleptonic to differential semileptonic
decay rates measured at $q^2=m^2_h$ is independent of
above-mentioned uncertainties. The determination of $|a_2^{\rm eff}|$ from
$B\to J/\psi K^{(*)}$ is sensitive to the
form factors $F_1^{BK}$, $A_{1,2}^{BK^*}$ and $V^{BK^*}$ at $q^2=m^2_{J/
\psi}$. In order to accommodate the observed production ratio
$R\equiv \Gamma(B\to J/\psi K^*)/\Gamma(B\to J/\psi K)$ by generalized
factorization, $a_2^{\rm eff}$ should be process insensitive; that is,
$a_2^{\rm eff}$
extracted from $J/\psi K$ and $J/\psi K^*$ final states should be very
similar. This puts a severe constraint on the form-factor models and
only a few models can satisfactorily explain the production ratio $R$.

The rest of this paper is organized as follows. In
Sec.~\ref{sec:bas}, we introduce the basic formula and the
classification of the relevant decay modes which have been
measured experimentally. Sec.~\ref{sec:FF} briefly describes various
form-factor models. The results and discussions for the effective parameters
$a_1^{\rm eff}$ and $a_2^{\rm eff}$ are presented in Secs.~IV and V,
respectively. Finally, the conclusion is given in Sec.~\ref{sec:con}.

\section{The basic framework}\label{sec:bas}
   Since, as we shall see below, the decays $B\to D^{(*)}D_s^{(*)},~J/\psi
K^{(*)}$ receive
penguin contributions, the relevant $\Delta B=1$ effective Hamiltonian
for our purposes has the form
\be \label{hamiltonian}
{\cal H}_{\rm eff} &=& {G_F\over\sqrt{2}}\Bigg\{ V_{cb}V_{uq}^*
\Big[c_1(\mu)O_1^{(uq)}(\mu)+c_2(\mu)O_2^{(uq)}(\mu)\Big]+V_{cb}V_{cs}^*
\Big[c_1(\mu)O_1^{(cs)}(\mu)+c_2(\mu)O_2^{(cs)}(\mu)\Big]  \non \\
&& -V_{tb}V_{ts}^*\sum^{10}_{i=3}c_i(\mu)O_i(\mu)\Bigg\}+{\rm h.c.},
\en
where
\be
&& O_1^{(uq)}= (\bar cb)_\vma(\bar qu)_\vma, \qquad\qquad\qquad\qquad
O_2^{(uq)} = (\bar qb)_\vma(\bar cu)_\vma, \non \\
&& O_1^{(cs)}= (\bar cb)_\vma(\bar sc)_\vma, \qquad\qquad\qquad\qquad~
O_2^{(cs)} = (\bar sb)_\vma(\bar cc)_\vma, \non \\
&& O_{3(5)}=(\bar qb)_\vma\sum_{q'}(\bar q'q')_{\vma(\vpa)}, \qquad  \qquad\,
O_{4(6)}=(\bar q_\alpha b_\beta)_\vma\sum_{q'}(\bar q'_\beta q'_\alpha)_{
\vma(\vpa)},   \\
&& O_{7(9)}={3\over 2}(\bar qb)_\vma\sum_{q'}e_{q'}(\bar q'q')_{\vpa(\vma)},
  \qquad\, O_{8(10)}={3\over 2}(\bar q_\alpha b_\beta)_\vma\sum_{q'}e_{q'}(\bar
q'_\beta q'_\alpha)_{\vpa(\vma)},   \non
\en
with $O_3$--$O_6$ being the QCD penguin operators and $O_{7}$--$O_{10}$
the electroweak penguin operators.

 To evaluate the decay amplitudes for the processes $B\to D^{(*)}D^{(*)}_s$,
$D^{(*)+,0}(\pi^-,\rho^-)$, $J/\psi K^{(*)}$, we first apply Eq.~(\ref{tree})
to the effective Hamiltonian (\ref{hamiltonian}) so that the factorization
approximation can be applied to the tree-level
hadronic matrix elements. We also introduce the shorthand notation
$X^{(BF_1,F_2)}$ to denote the factorized matrix element with the
$F_2$ meson being factored out~\cite{CT98}, for instance,
\be
X^{(B^-D^0,\pi^-)} &\equiv& \la \pi^-|(\bar du)_\vma|0\ra \la D^0|(\bar
cb)_\vma|B^-\ra, \non \\ X^{(B^-\pi^-,D^0)} &\equiv& \la D^0|(\bar
cu)_\vma|0\ra \la\pi^-|(\bar db)_\vma|B^-\ra\,.
\en
The results are:

\begin{itemize}
\item{Class I:} $\ov B_d^0 \to D^{(*)+} ~\pi^-(\rho^-)$\\
The decay amplitudes are given by
\be  \label{BI}
A(\ov B_d^0 \to D^{(*)+} \pi^-(\rho^-))=\frac{G_F}{\sqrt{2}}
\vcbud~ \left[a_1 X^{(\ov B^0D^{(*)+},\pi^-(\rho^-))}+a_2 X^{(\ov B^0,
D^{(*)+}\pi^-(\rho^-))}\right],
\en
where $X^{(\ov B, D^{(*)+}\pi^-(\rho^-))}$ is the factorized $W$-exchange
contribution.
\end{itemize}

\begin{itemize}
\item{Class I:} $B^- \to D^{(*)0}D_s^{(*)-}$ and  $\ov
B_d^0 \to D^{(*)+}~D_s^{(*)-}$\\
The decay amplitudes are given by
\be  \label{DD}
A(B \to D\,D_s)&=&\frac{G_F}{\sqrt{2}}\Bigg\{ \vcbcs~ a_1 -V_{tb}V_{ts}^*
\Big[ a_4+a_{10}  \non \\
&+& 2(a_6+a_8){m_{D_s}^2\over (m_b-m_c)(m_c+m_s)}\Big]\Bigg\}
X^{(BD,D_s)}  \non \\
&\cong& {G_F\over\sqrt{2}}\vcbcs \,\tilde a_1(B\to DD_s) X^{(BD,D_s)}
\,,
\en
where use of $V_{tb}V_{ts}^*\cong -\vcbcs$ has been made and
\be \label{DDa1}
\tilde a_1(B\to D\,D_s) = a_1\left( 1+{a_4+a_{10}\over a_1}+2{ a_6+a_8
\over a_1}\,{m_{D_s}^2\over (m_b-m_c)(m_s+m_c) }\right).
\en
Likewise,
\be  \label{DDa1'}
\tilde a_1(B\to D^*D_s) &=& a_1\left( 1+{a_4+a_{10}\over a_1}-2{ a_6+a_8
\over a_1}\,{m_{D_s}^2\over (m_b+m_c)(m_s+m_c) }\right),  \non \\
\tilde a_1(B\to D^{(*)}D_s^*) &=& a_1\left( 1+{a_4+a_{10}\over a_1}\right).
\en
Note that the decay $B^-\to D^0D_s^{(*)-}$ also receives a contribution from
the $W$-annihilation diagram, which is quark-mixing-angle doubly
suppressed, however.
\end{itemize}

\begin{itemize}
\item{Class II:} $\ov B_d^0 \to D^{(*)0} ~\pi^0(\rho^0)$\\
The factorized decay amplitudes are given by
\be \label{BII'}
A(\ov B_d^0 \to D^{(*)0} \pi^0(\rho^0))=\frac{G_F}{\sqrt{2}}
\vcbud~ a_2\left[X^{(\ov B^0\pi^0(\rho^0),D^{(*)0})}+ X^{(\ov B^0, D^{(*)0}
\pi^0(\rho^0))}\right],
\en
where $X^{(\ov B^0, D^{(*)0}\pi^0(\rho^0))}$ is the factorized
$W$-exchange contribution.
\end{itemize}

\begin{itemize}
\item{Class II:} $B^+ \to J/\psi~ K^{(*)+}$ and  $B^0 \to J/\psi~ K^{(*)0}$\\
The decay amplitudes are given by
\be  \label{BII} A(B \to J/\psi~
K^{(*)})=\frac{G_F}{\sqrt{2}}\vcbcs~ \tilde a_2 X^{(B K^{(*)},J/\psi)}\,,
\en
where
\be \label{a2}
\tilde a_2(B\to J/\psi K^{(*)}) \cong a_2\left[1+{a_3+a_5+a_7+a_9\over
a_2}\right].
\en
\end{itemize}

\begin{itemize}
\item{Class III:} $B^- \to D^{(*)0} ~\pi^-(\rho^-)$\\
The decay amplitudes are given by
\be \label{BIII}
A(B^- \to D^{(*)0} ~\pi^-(\rho^-))=\frac{G_F}{\sqrt{2}}\vcbud~ \Big[a_1
X^{(B^-D^{(*)0},\pi^-(\rho^-))}+ a_2
X^{(B^-\pi^-(\rho^-),D^{(*)0})}\Big]\,.
\en
\end{itemize}

\begin{itemize}
\item{Class III:} $B^- \to D^{(*)0}K^-$ \\
The factorized decay amplitudes are given by
\be  \label{BDK}
A(B^-\to D^{(*)0}K^-)=\frac{G_F}{\sqrt{2}}
V_{cb}V^*_{us}\left[a_1X^{(B^-D^{(*)0},K^-)+a_2 X^{(B^-K^-,D^{(*)0})}+}\right].
\en
\end{itemize}

\noindent Under the naive factorization approximation, $a_{2i}=
{c}_{2i}^{\rm eff}+{1\over N_c}{c}_{2i-1}^{\rm eff}$ and $a_{2i-1}=
{c}_{2i-1}^{\rm eff}+{1\over N_c}{c}^{\rm eff}_{2i}$ $(i=1,\cdots,5)$.
Since nonfactorizable effects can be absorbed into the parameters
$a_i^{\rm eff}$, this amounts to replacing $N_c$ in $a_i$
by $(N_c^{\rm eff})_i$ \cite{CT98} with
\be  \label{neff}
{1\over (\nc)_i}\equiv {1\over N_c}+\chi_i.
\en
Explicitly,
\be
a_{2i}^{\rm eff}={c}_{2i}^{\rm eff}+{1\over (N_c^{\rm eff})_{2i}}{c}_{2i-1}^{
\rm eff}, \qquad \quad a_{2i-1}^{\rm eff}=
{c}_{2i-1}^{\rm eff}+{1\over (N_c^{\rm eff})_{2i-1}}{c}^{\rm eff}_{2i}.
\en
(For simplicity, we have already dropped the superscript ``eff" of $a_i$
in Eqs.~(\ref{BI}) to (\ref{BDK}) and henceforth.)

   Although the purpose of the present paper is to treat the effective
coefficients $a_1$ and $a_2$ as free parameters to be extracted from
experiment, it is clear from Eqs.~(\ref{DDa1}) and (\ref{a2}) that the
determination of $a_{1,2}$ from
$B\to D^{(*)}D_s^{(*)}$ and $J/\psi K^{(*)}$ is contaminated by the
penguin effects. Therefore, it is necessary to make a theoretical
estimate on the penguin contribution. To do this, we employ the effective
renormalization-scheme and -scale independent Wilson coefficients
$c_i^{\rm eff}$ obtained at $k^2=m_b^2/2$ ($k$ being the gluon's
virtual momentum) \cite{CT98}:
\be   \label{ceff}
&& {c}_1^{\rm eff}=1.149, \qquad\qquad\qquad\qquad\qquad {c}_2^{\rm eff}=
-0.325,   \non \\
&& {c}_3^{\rm eff}=0.0211+i0.0045, \qquad\qquad\quad\,{c}_4^{\rm eff}=-0.0450
-i0.0136, \non  \\
&& {c}_5^{\rm eff}=0.0134+i0.0045, \qquad\qquad\quad\,{c}_6^{\rm eff}=-0.0560
-i0.0136, \non  \\
&& {c}_7^{\rm eff}=-(0.0276+i0.0369)\alpha, \qquad \quad {c}_8^{\rm eff}=
0.054\,\alpha, \non  \\
&& {c}_9^{\rm eff}=-(1.318+i0.0369)\alpha, \qquad \quad ~\,{c}_{10}^{\rm eff}=
0.263\,\alpha.
\en
For nonfactorizable effects, we choose $\nc(LL)\approx 2$ (see Sec.~V.E)
for $(V-A)(V-A)$
interactions (i.e. operators $O_{1,2,3,4,9,10}$) and $\nc(LR)\approx 5$
for $(V-A)(V+A)$ interactions (i.e. operators $O_{5,6,7,8}$). Our
choice for $\nc(LR)$ is motivated by the penguin-dominated charmless hadronic
$B$ decays (for details, see \cite{CT98,CCT}). Hence,
the theoretical values of the effective coefficients $a_i$ are given by
\be \label{ai}
&& a_1=0.986,  \qquad\qquad\qquad\qquad\qquad~~   a_2=0.25,   \non\\
&& a_3=-(0.00139+0.00226i),  \qquad \quad\, a_4=-(0.0344+0.0113i),  \non\\
&& a_5=0.0022+0.00181i, \qquad \qquad~~~\, a_6=-(0.0533+0.0127i), \non \\
&& a_7=-(1.24+2.73i)\times 10^{-4}, \qquad \quad a_8=(3.59-0.55i)\times
10^{-4},  \non \\
&& a_9=-(87.9+2.73i)\times 10^{-4}, \quad \qquad a_{10}=-(29.3+1.37i)\times
10^{-4}.
\en
From Eqs.~(\ref{DDa1}), (\ref{DDa1'}), (\ref{BII}) and (\ref{ai}), penguin
corrections to the tree amplitudes are found to be
\footnote{Our numerical estimate for the penguin effects in $B\to D\,D_s$
differs from \cite{Keum} due to different choices of $\nc(LL),~\nc(LR)$
and running quark masses.}
\be
&& |A_P/A_T|(B\to D\,D_s) = \left|{a_4+a_{10}\over a_1}+2{ a_6+a_8
\over a_1}\,{m_{D_s}^2\over (m_b-m_c)(m_s+m_c) }\right|=0.159, \non\\
&& |A_P/A_T|(B\to D^*D_s) = \left|{a_4+a_{10}\over a_1}-2{ a_6+a_8
\over a_1}\,{m_{D_s}^2\over (m_b+m_c)(m_s+m_c) }\right|=0.037, \non\\
&& |A_P/A_T|(B\to D^{(*)}D_s^*) = \left|{a_4+a_{10}\over a_1}
\right|=0.040, \non\\
&& |A_P/A_T|(B\to J/\psi K^{(*)}) = \left|{a_3+a_5+a_7+a_9\over a_2}\right|
=0.033\,,
\en
where we have used the current quark masses at the scale $\mu={\cal O}
(m_b)$: $m_s(m_b)=105\,{\rm MeV},~m_c(m_b)=0.95\,{\rm GeV},~m_b(m_b)=
4.34\,{\rm GeV}$.
Therefore, the penguin contribution to $B\to D^*D_s,~D^{(*)}D_s^*$ and
$J/\psi K^{(*)}$ is small, but its effect on $B\to D\,D_s$ is significant.
Numerically, the effective $\tilde a_i$ defined in Eqs.~(\ref{DDa1}) and
(\ref{a2}) are related to $a_i$ by
\be \label{penguin}
 \tilde a_1 (B\to D\,D_s) &=& 0.847\,a_1,  \non \\
 \tilde a_1 (B\to D^*D_s) &=& 1.037\,a_1,  \non \\
 \tilde a_1 (B\to D^{(*)}D_s^*) &=& 0.962\,a_1,  \non \\
 \tilde a_2 (B\to J/\psi K^{(*)}) &=& 0.968\,a_2.
\en

To evaluate the hadronic matrix elements,
we apply the following parametrization for decay constants and
form factors \cite{BSW85}
\be  \label{def}
\la 0|A_\mu|P(q)\ra &=&
if_Pq_\mu, \qquad \la 0|V_\mu|V(p,\vp)\ra=f_Vm_V\vp_ \mu,   \non
\\
\la P'(p')|V_\mu|P(p)\ra &=&
\left(p_\mu+p'_\mu-{m_P^2-m_{P'}^2\over q^2}\,q_ \mu\right)
F_1(q^2)+F_0(q^2)\,{m_P^2-m_{P'}^2\over q^2}q_\mu,   \non \\ \la
V(p',\vp)|V_\mu|P(p)\ra &=& {2\over
m_P+m_V}\,\epsilon_{\mu\nu\alpha \beta}\vp^{*\nu}p^\alpha p'^\beta
V(q^2),   \non \\
\la V(p',\vp)|A_\mu|P(p)\ra &=& i\Big[
(m_P+m_V)\vp_\mu A_1(q^2)-{\vp\cdot p\over m_P+m_V}\,(p+p')_\mu
A_2(q^2)    \non
\\ && -2m_V\,{\vp\cdot p\over
q^2}\,q_\mu\big[A_3(q^2)-A_0(q^2)\big]\Big],
\en
where $q=p-p'$, $F_1(0)=F_0(0)$, $A_3(0)=A_0(0)$,
\be
A_3(q^2)=\,{m_P+m_V\over 2m_V}\,A_1(q^2)-{m_P-m_V\over
2m_V}\,A_2(q^2)\,,
\en
and $P$, $V$ denote the pseudoscalar and vector mesons, respectively.
The factorized terms in (\ref{BI})-(\ref{BDK}) then have the expressions:
\be \label{term}
X^{(\ov B P_1,P_2)} &\equiv& \la P_2|
(\bar{q}_2 q_3)_\vma|0\ra\la P_1|(\bar {q}_1b)_\vma|\ov B
\ra=if_{P_2}(m_{B}^2-m^2_{P_1}) F_0^{ B P_1}(m_{P_2}^2),  \non \\
X^{(\ov BP,V)} &\equiv & \la V|
(\bar{q}_2 q_3)_\vma|0\ra\la P|(\bar{q}_1b)_\vma|\ov B
\ra=2f_V\,m_V F_1^{ B P}(m_{V}^2)(\vp^*\cdot p_{_{B}}),   \non \\
X^{(\ov BV,P)} &\equiv & \la P |
(\bar{q}_2 q_3)_\vma|0\ra\la V|(\bar{q}_1b)_\vma|\ov B
\ra=2f_P\,m_V A_0^{ B V}(m_{P}^2)(\vp^*\cdot p_{_{B}}),  \non \\
X^{(\ov BV_1,V_2)} &\equiv & \la V_2 |
(\bar{q}_2 q_3)_\vma|0\ra\la V_1|(\bar{q}_1b)_\vma|\ov B \ra
=- if_{V_2}m_{_{2}}\Bigg[ (\vp^*_1\cdot\vp^*_2)
(m_{B}+m_{1})A_1^{ BV_1}(m_{2}^2)  \non \\
&-& (\vp^*_1\cdot p_{_{B}})(\vp^*_2
\cdot p_{_{B}}){2A_2^{ BV_1}(m_{2}^2)\over (m_{B}+m_{1}) }
+ i\epsilon_{\mu\nu\alpha\beta}\vp^{*\mu}_2\vp^{*\nu}_1p^\alpha_{_{B}}
p^\beta_1\,{2V^{ BV_1}(m_{2}^2)\over (m_{B}+m_{1}) }\Bigg],
\en
where $\vp^*$ is the polarization vector of the vector meson $V$.

With the factorized decay amplitudes given in Eqs.~(\ref{BI})-(\ref{BDK}),
the decay rates for $B\to PP,~VP$ are given by
\be
\Gamma(B\to P_1P_2) &=& \frac{ p_c}{8\pi m_B^2}|A(B\to P_1P_2)|^2 \,, \non\\
\Gamma(B\to VP) &=& {p_c^3\over 8\pi m^2_V} |A(B\to VP)/(\vp\cdot
p_{_{B}})|^2,
\en
where
\be
p_c=\frac{\sqrt{[m_B^2-(m_1+m_2)^2][m_B^2-(m_1-m_2)^2]}}{2m_B}
\en
is the c.m. momentum of the decay particles.
For simplicity, we consider a single factorizable amplitude for $B\to VV$:
$A(B\to V_1V_2)=\alpha X^{(BV_1,V_2)}$. Then
\be
\Gamma(B\to V_1V_2)={p_c\over 8\pi m^2_{_{B}} }|\alpha(m_B+m_1)m_2 f_{V_2}
A_1^{BV_1}(m^2_2)|^2H,
\en
with
\be \label{H}
H = (a-bx)^2+2(1+c^2y^2),
\en
and
\be \label{abc}
&&  a={m_{B}^2-m_1^2-m_2^2\over 2m_1m_2}, \qquad b={2m^2_{B}p_c^2\over
m_1m_2(m_{B}+m_1)^2}, \qquad c={2m_{B}p_c\over (m_{B}+m_1)^2}, \non\\
&& x={A_2^{BV_1}(m_2^2)\over A_1^{BV_1}(m^2_2)}, \qquad \quad \qquad
y={V^{BV_1}(m_2^2)\over A_1^{BV_1}(m_2^2)},
\en
where $m_1$ ($m_2$) is the mass of the vector meson $V_1$ ($V_2$).

  From Eqs.~(\ref{BI}-\ref{BIII}) we see that $|a_1|$ can be determined from
$\ov B^0\to D^{(*)+}(\pi^-,\rho^-),~B\to D^{(*)}D_s^{(*)}$, $|a_2|$
from $B\to J/\psi K^{(*)},~\ov B^0\to D^{(*)0}(\pi^0,\rho^0)$, provided that
the $W$-exchange contribution is negligible in $B\to D^{(*)}(\pi,\rho)$ decays
and that penguin corrections are taken into account. It is also clear that
the ratio $a_2/a_1$ can be determined from the ratios of charged to neutral
branching fractions:
\be  \label{R14}
R_1\equiv{{\cal B}(B^-\to D^0\pi^-)\over{\cal
B}(\ov{B}^0\to D^+\pi^-)} &=& \r\left(1+{m_B^2-m_\pi^2\over
m_B^2-m_D^2}{f_D\over f_\pi}{F_0^{B\pi}(m_D^2) \over
F_0^{BD}(m_\pi^2)}{a_2\over a_1}\right)^2,   \non \\
R_2\equiv{{\cal B}(B^-\to D^0\rho^-)\over{\cal B}(\ov{B}^0\to
D^+\rho^-)} &=& \r\left(1+{f_D\over f_\rho}{A_0^{B\rho}(m_D^2)
\over F_1^{BD}(m_\rho^2)}{a_2\over a_1}\right)^2,   \non \\
R_3\equiv{{\cal B}(B^-\to D^{*0}\pi^-)\over{\cal B}(\ov{B}^0\to
D^{*+}\pi^-)} &=& \r\left(1+{f_{D^*}\over
f_\pi}{F_1^{B\pi}(m_{D^*}^2) \over A_0^{BD^*}(m_\pi^2)}{a_2\over
a_1}\right)^2,    \\
R_4\equiv{{\cal B}(B^-\to
D^{*0}\rho^-)\over{\cal B}(\ov{B}^0\to D^{*+} \rho^-)} &=&
\r\left(1+2\eta{H_1\over H}+\eta^2{H_2\over H}\right),   \non
\en
with
\be
\eta &=& {m_{D^*}(m_B+m_\rho)\over
m_\rho(m_B+m_{D^*})}\,{f_{D^*}\over f_\rho}
\,{A_1^{B\rho}(m^2_{D^*})\over A_1^{BD^*}(m^2_\rho)}\,{a_2\over
a_1},  \non \\
H_1 &=&
(a-bx)(a-b'x')+2(1+cc'yy'),   \\ H_2 &=& (a-b'x')^2+2(1+c'^2y'^2),
\non
\en
where $H,a,b,c,x,y$ are those defined in Eqs.~(\ref{H}) and (\ref{abc})
with $V_1=D^*$ and
$V_2=\rho$, and $b',c',x',y'$ are obtained from $b,~c,~x,~y$ respectively
with the replacement $D^*\leftrightarrow \rho$.

\section{Model calculations of form factors}\label{sec:FF}
The analyses of $a_2$, $a_1$, and $a_2/a_1$ depend strongly on
the form factors chosen for calculations.
In the following study, we will consider six distinct form-factor models:
the Bauer-Stech-Wirbel (BSW) model \cite{BSW85,bsw}, the modified BSW
model (referred to as the NRSX model) \cite{NRSX}, the relativistic
light-front (LF) quark model \cite{cch},
the Neubert-Stech (NS) model \cite{ns}, the QCD sum rule calculation by Yang
\cite{yang}, and the light-cone sum rule (LCSR) analysis
\cite{bb1}.

    Form factors in the BSW model are calculated at zero momentum transfer
in terms of relativistic bound-state wave functions obtained in the
relativistic harmonic oscillator potential model \cite{BSW85}. The
form factors at other values of $q^2$ are obtained from that at $q^2=0$ via
the pole dominance ansatz
\be \label{q2}
F(q^2)=\frac{F(0)}{(1-q^2/m_{pole}^2)^n},
\en
where $m_{pole}$ is the appropriate pole mass.
The BSW model assumes a monopole behavior (i.e. $n=1$) for all the form
factors. However, this is not consistent with heavy quark symmetry for
heavy-to-heavy transition. In the heavy quark limit, the $B\to D$ and
$B\to D^*$ form factors are all related to a single Isgur-Wise function
through the relations
\be \label{hqs}
{m_B+m_D\over 2\sqrt{m_Bm_D} }\,\xi(v_B\cdot v_D) &=& F_1^{BD}(q^2)=
{F_0^{BD}(q^2)\over 1-q^2/(m_B+m_D)^2},   \non \\
{m_B+m_{D^*}\over 2\sqrt{m_Bm_{D^*}} }\,\xi(v_B\cdot v_{D^*}) &=&
V^{BD^*}(q^2)=A_0^{BD^*}(q^2)=A_2^{BD^*}(q^2)   \non \\
&=& {A_1^{BD^*}(q^2)\over 1-q^2/(m_B+m_{D^*})^2}.
\en
Therefore, the form factors $F_1,V,A_0,A_2$ in the infinite quark mass
limit have the same $q^2$ dependence and they differ from $F_0$ and $A_1$
by an additional pole factor. In general, the heavy-to-heavy form factors
can be parametrized as
\be \label{HQET}
F^{BD}_0(q^2)&&=\left(\BD\right)^{-1} {\omega_D(q^2)+1\over
2}\,{1\over r(q^2)}\G(q^2)\,,\non\\
F^{BD}_1(q^2)&&=\left(\BD\right)
\G(q^2)\,,\non\\
A^{BD^*}_0(q^2)&&=\left(\frac{m_B+m_{D^*}}{2m_{D^*}}A^{BD^*}_1(0)-
\frac{m_B-m_{D^*}}{2m_{D^*}}A^{BD^*}_2(0)\right)\frac{\F'(q^2)}
{\F'(0)}\,,   \non \\
A^{BD^*}_1(q^2)&&=\left(\BDstar\right)^{-1}
{\omega_{D^*}(q^2)+1\over 2}\F(q^2)\,,\non\\
A^{BD^*}_2(q^2)&&=\left(\BDstar\right) \F(q^2)~r_2(q^2)\,, \non\\
V^{BD^*}(q^2)&&=\left(\BDstar\right) \F(q^2)~r_1(q^2)\,,
\en
where
\be
\omega_{D^{(*)}}(q^2) && \equiv v_B\cdot v_{D^{(*)}}
=\frac{m_B^2+m_{D^{(*)}}^2-q^2}{2m_B m_{D^{(*)}}}\,,\non\\
\G(q^2)&&= \G(q_{max}^2)[1-\rho_D^2(\omega_D(q^2)-1)]\,,\non\\
\F(q^2)&&=
\F(q_{max}^2)[1-\rho_{D^*}^2(\omega_{D^{(*)}}(q^2)-1)]\,,\non\\
\F'(q^2)&&=
\F'(q_{max}^2)[1-\rho'^2(\omega'(q^2)-1)]\,,\non\\
r(q^2) && =\left[1-\frac{q^2}{(m_B+m_{D})^2}\right]\frac{
F_1^{BD}(q^2)}{F^{BD}_0(q^2)}\,,\non\\
r_1(q^2)&&=\left[1-\frac{q^2}{(m_B+m_{D^*})^2}\right]\frac{
V^{BD^*}(q^2)}{A^{BD^*}_1(q^2)}\,,\non\\
r_2(q^2)&&=\left[1-\frac{q^2}{(m_B+m_{D^*})^2}\right]\frac{
A_2^{BD^*}(q^2)}{A^{BD^*}_1(q^2)}\,.
\en
In the heavy quark limit $m_b\to\infty$, the two form factors $\F(q^2)$ and
$\G(q^2)$, whose slopes are  $\rho^2_{D^{(*)}}$, coincide with the
Isgur-Wise function $\xi(q^2)$, and $r(q^2),r_1(q^2)$ as well as
$r_2(q^2)$ are equal to unity. The $q^2$ dependence of $B\to D^{(*)}$
form factors in the NRSX and NS models is more complicated because
perturbative
hard gluon and nonperturbative $1/m_Q$ corrections to each form factor
are taken into consideration and moreover these corrections by themselves
are also $q^2$ dependent (see \cite{NRSX} for more details).

   Form factors for heavy-to-heavy and heavy-to-light transitions at
time-like momentum transfer are explicitly calculated in the LF model.
It is found in \cite{cch} that the form factors $F_1,V,A_0,A_2$ all
exhibit a dipole behavior, while $F_0$ and $A_1$ show a monopole dependence
in the close vicinity of maximum
recoil (i.e. $q^2=0$) for heavy-to-light transitions and in a broader
kinematic region for heavy-to-heavy decays. Therefore, the $q^2$ dependence
of $B\to D^{(*)}$ form factors in the heavy quark limit
is consistent with the requirement of heavy quark symmetry.
Note that the pole mass in this model obtained by fitting the calculated
form factors to Eq.~(\ref{q2}) is slightly
different from that used in the BSW model (see Table I).

  Due to the lack of analogous heavy quark symmetry, the calculation of
heavy-to-light transitions is rather model dependent. In addition to
the above-mentioned BSW and LF models, form factors for the $B$ meson
to a light meson are also considered in many other models. The NRSX model
takes the BSW model results for the form factors at zero momentum
transfer but makes a different ansatz for their $q^2$ dependence, namely
a dipole behavior (i.e. $n=2$) is assumed for the
form factors $F_1,~A_0,~A_2,~V$, motivated by the heavy-quark-symmetry
relations (\ref{hqs}), and a monopole dependence for $F_0,A_1$.
The heavy-to-light form factors in the NS model
have the expressions \cite{ns}:
\be
F^{BP}_0(q^2)&&=\left(\BP\right)^{-1}\ccPin~
\frac{1}{1+r_V\frac{\omega_{BP}(q^2)-1}{\omega_{BP}(0)-1}}\,,\non\\
F^{BP}_1(q^2)&&=\BP\ccP~ \frac{1}{1+r_V}~
\frac{\omega_{BP}(0)-\omega_{BP}(m_{1^-}^2)}{\omega_{BP}(q^2)-\omega_{BP}
(m_{1^-}^2)}\,,\non\\
A^{BV}_0(q^2)&&=\BV\ccV~ \frac{1}{1+r_V}~
\frac{\omega_{BV}(0)-\omega_{BV}(m_{0^-}^2)}{\omega_{BV}(q^2)-\omega_{BV}
(m_{0^-}^2)}\,,\non\\
A^{BV}_1(q^2)&&=\left(\BV\right)^{-1}\ccVin~\frac{1}{1+r_V\frac{\omega_{BV}
(q^2)-1}{\omega_{BV}(0)-1}}\,,\non\\
A^{BV}_2(q^2)&&=\BV\ccV~ \frac{1}{1+r_V}~\frac{\omega_{BV}(0)-\omega_{BV}
(m_{1^+}^2)}{\omega_{BV}(q^2)-\omega_{BV}(m_{1^+}^2)}
\,,\non\\
V^{BV}(q^2)&&=\BV\ccV~ \frac{1}{1+r_V}~
\frac{\omega_{BV}(0)-\omega_{BV}(m_{1^-}^2)}{\omega_{BV}(q^2)-\omega_{BV}
(m_{1^-}^2)}\,,
\en
where
\be
\omega_{BP(V)}&&=\frac{m_B^2+m_{BP(V)}^2-q^2}{2m_B
m_{BP(V)}}\,,\non\\ r_V&&=\frac{(m_B-m_V)^2}{4m_B
m_V}\left(1+\frac{4m_B m_V}{m_{1^+}^2-(m_B-m_V)^2}\right)\,.
\en
Here $m_{0^-}, m_{1^-}$, and $m_{1^+}$ are the lowest resonance
states with the quantum numbers $0^-$, $1^-$, and $1^+$,
respectively.\footnote{Following \cite{ns}, we will simply add
400~MeV to $m_{1^-}$ to obtain the masses of $1^+$
resonances.}

   We consider two QCD sum rule calculations for $B$-to-light
transitions. The form factors $F_1$ and $A_1$
in the Yang's sum rule have a monopole behavior, while $A_2$ and $V$
show a dipole $q^2$ dependence. The momentum
dependence of the form factors $F^{B\pi(K)}_0$ and $A^{B\rho(K^*)}_0$ is
slightly complicated and is given by \cite{yang}
\be
F^{B\pi}_0(q^2)&&=-0.28\left(\frac{5.4^2}{5.4^2-q^2}\right)
\frac{q^2}{m_B^2-m_\pi^2} +F^{B\pi}_1(q^2)\,,\non\\
F^{BK}_0(q^2)&&=-0.32\left(\frac{5.8^2}{5.8^2-q^2}\right)
\frac{q^2}{m_B^2-m_K^2} +F^{BK}_1(q^2)\,,\non\\
A^{B\rho}_0(q^2)&&=0.015
q^2\left(\frac{5.98^2}{5.98^2-q^2}\right)^2
+\left(\frac{m_B+m_\rho}{2m_\rho}A^{B\rho}_1(q^2)-
\frac{m_B-m_\rho}{2m_\rho}A^{B\rho}_2(q^2)\right)\,,\non\\
A^{BK^*}_0(q^2)&&=0.02 q^2\left(\frac{6.1^2}{6.1^2-q^2}\right)^2
+\left(\frac{m_B+m_{K^*}}{2m_K{^*}}A^{BK{^*}}_1(q^2)-
\frac{m_B-m_K{^*}}{2m_K{^*}}A^{BK{^*}}_2(q^2)\right)\,.
\en
The $q^2$ behavior of $B$-to-light form factors in the LCSR analysis of
\cite{bb1} are parametrized as
\be   \label{lc}
F(q^2)=\frac{F(0)}{1-a_F{q^2\over
m_B^2}+b_F({q^2\over m_B^2})^2}\,,
\en
where the relevant fitted parameters $a_F$ and $b_F$ can be found in
\cite{bb1}.

 Since only the form factors for $B$-to-light transition are evaluated in
the Yang's sum rule analysis and the LCSR, we shall
adopt the parametrization (\ref{HQET}) for the $B \to D^{(*)}$ form
factors, in which the relevant parameters are chosen in such a way that
$B\to D^{(*)}$ transitions in the NS model are reproduced:
\be \label{srneubert}
&& \F(q^2_{max})=0.88\,,  \qquad \quad
\G(q^2_{max})=1.00 \,,\nonumber\\
&& \rho^2_{D}=0.62\,, \qquad \rho'^2=0.62\,, \qquad
\rho_{D^*}^2= 0.91\,,  \non\\
&& r(q^2) \approx 1\,,  \qquad
r_1(q^2)\approx
r_1=1.3\pm 0.1\,,  \qquad
r_2(q^2)\approx r_2=0.8\pm 0.2\,,
\en
as a supplement to the Yang's~\cite{yang} and LCSR~\cite{bb1}
calculations. The theoretical prediction for $r_1$ and $r_2$ \cite{neubert}
is in good agreement with the CLEO measurement \cite{rb}: $r_1=1.18\pm 0.32$
and $r_2=0.71\pm 0.23$ obtained at zero recoil. Note that the
predictions of $B\to D^{(*)}$ form factors are slightly different in the
NRSX and NS models (see Table I) presumably due to the use of different
Isgur-Wise functions.

   To close this section, all the form factors relevant to the present paper
at zero momentum
transfer in various models and the pole masses available in the BSW and LF
models and in the Yang's sum rules are summarized in Table I.

\section{Determination of \lowercase{$a_1$} }\label{sec:res}

   In order to extract the effective coefficient $a_1$ from $\ov B^0\to
D^{(*)+}(\pi^-,\rho^-)$ and $B\to D^{(*)}D_s^{(*)}$ decays, it is necessary
to make several assumptions: (i) the $W$-exchange contribution in
$\ov B^0\to D^{(*)}\pi(\rho)$ is negligible, (ii) penguin corrections
can be reliably estimated,
and (iii) final-state interactions can be neglected. It is known that
$W$-exchange is subject to helicity and color
suppression, and the helicity mismatch is expected to be more effective
in $B$ decays because of the large mass of the $B$ meson.
Final-state interactions for $B\to D^{(*)}(\pi,\rho)$ decays
are customarily parametrized in terms of isospin phase shifts for
isospin amplitudes. Intuitively, the phase shift difference $\delta_{1/2}-
\delta_{3/2}$, which is of order $90^\circ$ for $D\to \ov K\pi$ modes, is
expected to play a much minor role in the energetic $B\to D\pi$ decay, the
counterpart of $D\to\ov K\pi$ in the $B$ system, as the decay particles
are moving fast, not allowing adequate time for final-state interactions.
From the current CLEO limit on $\ov B^0\to D^0\pi^0$ \cite{Nemati}, we find
\cite{ns}
\be
\sin^2(\delta_{1/2}-\delta_{3/2})\leq {9\over 2}\,{\tau(B^-)\over
\tau(\ov B^0)}\,{{\cal B}(\ov B^0\to D^0\pi^0)\over {\cal B}(B^-\to D^0\pi^-)}
=0.109\,,
\en
and hence
\be  \label{19}
|\delta_{1/2}-\delta_{3/2}|_{B\to D\pi}<19^\circ.
\en
We shall see in Sec.~V.III and in Fig.~1 that the effect of final-state
interactions \footnote{Final-state interactions usually vary from channel
to channel.
For example, $|\delta_{1/2}-\delta_{3/2}|$ is of order $90^\circ$ for $D\to
\ov K\pi,~\ov K^*\pi$, but it is consistent with zero isospin phase shift for
$D\to \ov K\rho$. The preliminary CLEO studies of the helicity amplitudes
for the decays $\ov B^0\to D^{*+}\rho^-$ and $B^-\to D^{*0}\rho^-$ indicate
some non-trivial phases which could be due to FSI \cite{Bon}. At any
rate, FSI are expected to be important for the determination of the
effective coefficient $a_2$ (see Sec.~V.III), but not for $a_1$.}
subject to the above phase-shift constraint is negligible on
$\Gamma(\ov B^0\to D^+\pi^-)$ and hence
it is justified to neglect final-state interactions for determining $a_1$.
The extraction of $a_2$ from $B\to J/\psi K^{(*)}$ does not suffer from
the above ambiguities (i) and (iii). First, $W$-exchange does not contribute
to this decay mode. Second, the $J/\psi K^{(*)}$ channel is a single
isospin state.

\subsection{Model-dependent extraction}
  We will first extract $a_1$ from the data in a model-dependent manner and
then come to an essentially model-independent method for determining the
same parameter.

  Armed with the form factors evaluated in various models for $B\to D$ and
$B\to D^*$ transitions, we are ready to determine the effective coefficient
$a_1$ from the data of $\ov B^0\to D^{(*)+}(\pi^-,\rho^-)$ and
$B\to D^{(*)}D_s^{(*)}$ decays \cite{PDG}. The results are shown in
Tables II and
III in which we have taken into account penguin corrections to $a_1$ [see
Eq.~(\ref{penguin})]. We will choose the sign
convention in such a way that $a_1$ is positive; theoretically, it
is expected that the sign of $a_1$ is the same as $c_1$.
In the numerical analysis, we adopt the following parameters, quark-mixing
matrix elements: $|V_{cb}|=0.039\pm 0.003, |V_{ud}|=|V_{cs}|=0.975\pm 0.001$;
decay constants: $f_\pi$=132 MeV, $f_K=160$ MeV, $f_{\rho}$=216 MeV,
$f_D$=200 MeV,
$f_{D^*}$=230 MeV, $f_{D_s}$=240 MeV, $f_{D_s^*}$=275 MeV,
$f_{J/\psi}$=394 MeV, and lifetimes: $\tau(\ov B^0)=(1.57\pm 0.03)~ps$,
$\tau(B^-)=(1.67\pm 0.03)~ps$ \cite{LEP}.
Because of the
uncertainties associated with the decay constants $f_{D_s}$ and $f_{D_s^*}$,
the value of $a_1$ obtained from $B\to D^{(*)}D_s^{(*)}$ decays in Table III
is normalized at $f_{D_s}=240$ MeV and $f_{D_s^*}=275$ MeV.
For example, $a_1$ determined in the NRSX model reads
\be
&& a_1\left(\ov B^0\to D^{(*)+}(\pi^-,\rho^-)\right) = 1.04\pm 0.03\pm 0.08,
\non \\
&& a_1\left( B\to D^{(*)}D_s\right) = (1.26\pm 0.11\pm 0.09)\times \left(
{240\,{\rm MeV}\over f_{D_s}}\right),  \non \\
&& a_1\left( B\to D^{(*)}D_s^{*}\right) = (1.12\pm 0.12\pm 0.08)\times \left(
{275\,{\rm MeV}\over f_{D^*_s}}\right),
\en
where the first error comes from the experimental branching ratios and
the second one from the $B$ meson lifetimes and quark-mixing matrix elements.
Evidently, $a_1$ lies in the vicinity of unity.

  Several remarks are in order. (i) From Tables II and III we see
that $a_1$ extracted from $B\to D^{(*)}D_s^{(*)}$ is consistent with that
determined from $B\to D^{(*)}\pi(\rho)$, though its central value
is slightly larger in the former. (ii) Theoretically, it is expected that
$\Gamma(\ov B^0\to D^{(*)+}D_s^{(*)-})=\Gamma(B^-\to D^{(*)0}D_s^{(*)-})$
and hence ${\cal B}(B^-\to D^{(*)0}D_s^{(*)-})\approx 1.07\,{\cal B}
(B^0\to D^{(*)+}D_s^{(*)-})$. The errors of the present data are too large
to test this prediction. (iii) The central value of $a_1$ extracted from
$\ov B^0\to D^{*+}D_s^-$ and $B^-\to D^{*0}
D_s^-$ in the BSW model deviates substantially from unity.
This can be understood as follows. The decay amplitude of the above two modes
is governed by the form factor $A_0^{BD^*}(m^2_{D_s})$. However, the
$q^2$ dependence of $A_0(q^2)$ in this model is of the monopole form so
that $A_0$ does not increase with $q^2$ fast enough compared to the other
form-factor models.

\subsection{Model-independent or model-insensitive extraction}

   As first pointed out by Bjorken \cite{Bjorken}, the decay rates of
class-I modes can be related under the factorization hypothesis to the
differential semileptonic decay widths at the appropriate $q^2$.
More precisely,
\be   \label{S}
S_h^{(*)}\equiv {{\cal B}(\ov B^0\to D^{(*)+}h^-)\over d{\cal B}(\ov B^0\to
D^{(*)+}\ell^-\bar \nu)/dq^2\Big|_{q^2=m^2_h} }=6\pi^2\tilde a_1^2f_h^2
|V_{ij}|^2Y_h^{(*)},
\en
where $\tilde a_1=a_1$ in the absence of penguin corrections [the
expressions of $\tilde a_1$ are given in Eq.~(\ref{DDa1})],
$V_{ij}=V_{ud}$ for $h=\pi,\rho$, $V_{ij}=V_{cs}$ for $h=D_s^{(*)}$,
and \cite{NRSX}
\be   \label{Y}
 Y_P &=& { (m_B^2-m_D^2)^2\over [m_B^2-(m_D+m_P)^2][m_B^2-(m_D-m_P)^2] }\,
\left| { F_0^{BD}(m_P^2)\over F_1^{BD}(m_P^2) }\right|^2,  \non \\
 Y^*_P &=& { [m_B^2-(m_{D^*}+m_P)^2][m_B^2-(m_{D^*}-m_P)^2]\over m_P^2}\,{
\left| A_0^{BD^*}(m_P^2)\right|^2\over \sum_ {i=0,\pm 1}\left| H_i^{BD^*}
(m_P^2)\right|^2 },  \non \\
  Y_V &=& Y_V^*=\,1\,,
\en
with the helicity amplitudes $H_0(q^2)$ and $H_\pm (q^2)$  given by
\be
H_\pm^{BD^*}(q^2) &=& (m_B+m_{D^*})A_1^{BD^*}(q^2)\mp {2m_Bp_c\over m_B+m_{D^*}
 }\,V^{BD^*}(q^2),  \non \\
H_0^{BD^*}(q^2) &=& {1\over 2m_{D^*}\sqrt{q^2}} \Bigg[ (m_B^2-m^2_{D^*}-q^2)
(m_B+m_{D^*})A_1^{BD^*}(q^2)   \non \\
&& -{4m_B^2p_c^2\over m_B+m_{D^*} } A_2^{BD^*}(q^2)\Bigg],
\en
where $p_c$ is the c.m. momentum.

Since the ratio $S_h^{(*)}$ is independent of $V_{cb}$ and form factors,
its experimental measurement can be utilized to fix $a_1$ in a
model-independent manner, provided that $Y_h^{(*)}$ is also independent
of form-factor models. From Table IV we see that $Y_\pi^*$ and in particular
$Y_\pi$ are essentially model independent. The BSW model has a larger
value for $Y_{D_s}$ and a smaller value for $Y_{D_s}^*$ compared to the
other models because all the form factors in the former are
assumed to have the same monopole $q^2$ behavior, a hypothesis not in
accordance with heavy quark symmetry. In the heavy quark limit, one has
$Y_{D_s}\approx 1.36$ and $Y_{D_s}^*\approx 0.37$  \cite{NRSX};
the former is quite close to the model calculations (see Table IV).
In short, $Y_\rho^{(*)},~Y_{D_s^*}^{(*)},~Y_\pi$ are model independent,
$Y_\pi^*$ is model insensitive, while $Y_{D_s}$ and $Y_{D_s}^*$ show a
slight model dependence.

  In Table V  the experimental data of $d{\cal B}(\ov B^0\to D^+\ell^-\bar\nu)
/dq^2$ (at $q^2=m_\pi^2$ and $m_\rho^2$) and $d{\cal B}(\ov B^0\to D^{*+}
\ell^-\bar\nu)/dq^2$ are taken from \cite{Browder} and \cite{Bergfeld},
respectively. Note that the ``data" of $d{\cal B}(\ov B^0\to D^+\ell^-\bar
\nu)/dq^2$ at small $q^2$ are actually obtained by first performing a fit
to the experimental differential $q^2$ distribution and then interpolating
it to $q^2=m_\pi^2$ and $m_\rho^2$. For the data of
$d{\cal B}(\ov B^0\to D^+\ell^-\bar\nu)/dq^2$ at $q^2=m^2_{D_s}$ and
$m^2_{D_s^*}$ we shall use the CLEO data for $d\Gamma/d\omega$
expressed in the form \cite{Artuso}
\be \label{diff}
{d\Gamma(B\to D\ell\bar\nu)\over d\omega} =\,{G_F^2\over 48\pi^2}\,
(m_B+m_D)^2m_D^3(\omega^2-1)^{3/2}|V_{cb}\F(\omega)|^2,
\en
where $\omega\equiv v_B\cdot v_D=(m_B^2+m_D^2-q^2)/(2m_Bm_D)$. A fit
of $\F(\omega)$ parametrized in the linear form
\be
\F(\omega)=\F(1)[1-\rho^2(\omega-1)],
\en
to the CLEO data yields \cite{Artuso}
\be \label{slope}
\rho^2=0.81\pm 0.14,  \qquad |V_{cb}\F(1)|=(4.31\pm 0.42)\times 10^{-2}.
\en
From (\ref{diff})-(\ref{slope}) we obtain $d{\cal B}(\ov B^0\to D^+\ell^-
\bar\nu)/dq^2$ at $q^2=m^2_{D_s}$ and $m^2_{D_s^*}$ as shown in Table V.
Note that we have applied the relation ${\cal B}(B^-\to D^{(*)0}D_s^{(*)-})
\approx 1.07\,{\cal B}(B^0\to D^{(*)+}D_s^{(*)-})$ to get the average
branching ratio for $B\to D^{(*)}D_s^{(*)}$ and the ratios $S^{(*)}_{D_s}$
and $S^{(*)}_{D_s^*}$.
It is easy to check that the data, say $d{\cal B}/dq^2=(0.35\pm 0.06)\times
10^{-2}\,{\rm GeV}^{-2}$ at $q^2=m_\pi^2$, are well reproduced through this
interpolation.

   The results of $a_1$ extracted in this model-independent or
model-insensitive way are exhibited in Table V (for a recent similar work,
see \cite{Ciuchini}), where we have chosen
$Y_{D_s}=1.36$ and $Y^*_{D_s}=0.40$ as representative values.
As before, the value of $a_1$ obtained from $B\to D^{(*)}D_s^{(*)}$ decays
is normalized at $f_{D_s}=240$ MeV and $f_{D_s^*}=275$ MeV.
In view of the present theoretical and experimental uncertainties with
the decay constants $f_{D_s}$ and $f_{D_s^*}$ and the relatively small
errors with the data of $D\pi$ and $D^*\pi$ final states, we believe that
the results (see Table V)
\be
a_1(\ov B^0\to D^+\pi^-)=0.93\pm 0.10, \qquad \quad a_1(\ov B^0\to D^{*+}
\pi^-)=1.09\pm 0.07
\en
are most reliable and trustworthy. Of course, if the factorization hypothesis
is exact, $a_1$ should be universal and process independent. However, we
have to await more precise measurement of the differential distribution
in order to improve the values of $a_1$ and to
have a stringent test on factorization.

Once $a_1$ is extracted from $S_h^{(*)}$, some of the $B\to D^{(*)}$ form
factors can be determined from the measured $B\to D^{(*)}(\pi,\rho)$ and
$D^{(*)}D_s^{(*)}$ rates in a model-independent way:
\be  \label{form}
F_0^{BD}(m_\pi^2) =0.66\pm 0.06\pm 0.05\,, && \qquad\quad F_0^{BD}(m^2_{D_s})
=0.78\pm 0.08\pm 0.06\,,    \non \\
F_1^{BD}(m_\rho^2) = 0.67 \pm 0.06\pm 0.05\,, && \qquad\quad F_1^{BD}(m^2_{
D_s^*})= 0.89\pm 0.10\pm 0.07\,,   \non \\
A_0^{BD^*}(m_\pi^2)=0.56\pm 0.03\pm 0.04\,, && \qquad\quad A_0^{BD^*}(m^2_{
D_s})=0.77\pm 0.03\pm 0.06\,.
\en
It should be stressed that the above form-factor extraction is independent
of the decay constants $f_{D_s}$ and $f_{D_s^*}$.
It is interesting to see that $F_1^{BD}$ tends to increase with $q^2$ faster
than $F_0^{BD}$, in agreement with the heavy-quark-symmetry requirement
(\ref{hqs}).

  The decay constants $f_{D_s}$ and $f_{D_s^*}$ can be extracted if $a_1$
determined from $B\to D^{(*)}D_s^{(*)}$ is assumed to be the same as
that from $D^{(*)}\pi(\rho)$ channels. For example, the assumption
of $a_1(\ov B^0\to D^+ D_s^-)=a_1(\ov B^0\to D^+\pi^-)$ will lead to an
{\it essentially model-independent} determination of $f_{D_s}$. We see
from Table V that
\be
(1.01\pm 0.14)(240\,{\rm MeV}/f_{D_s})=\,0.93\pm 0.10\,,
\en
and hence
\be
f_{D_s}=(261\pm 46)\,{\rm MeV}.
\en
Another equivalent way of fixing $f_{D_s}$ is to consider the ratio of
hadronic decay rates \cite{ns}
\be   \label{rr}
{ {\cal B}(\ov B^0\to D^+ D_s^-)\over {\cal B}(\ov B^0\to D^+ \pi^-) }=
\left(0.847\,{F_0^{BD}(m^2_{D_s})\over F_0^{BD}(m^2_{\pi})}\,{f_{D_s}\over
f_\pi }\right)^2{1.812\,{\rm GeV}\over 2.306\,{\rm GeV} },
\en
where 1.812 GeV and 2.306 GeV are the c.m. momenta of the decay particles
$D_s$ and $\pi$, respectively, use of $a_1(B\to DD_s)=a_1(B\to D\pi)$ has
been made and penguin corrections have been included.
It is easy to check that the same value of $f_{D_s}$ is
obtained when the model-independent form factors (\ref{form}) are applied to
(\ref{rr}). Likewise,
\be
f_{D_s^*}=(266\pm 62)\,{\rm MeV}
\en
is obtained by demanding $a_1(\ov B^0\to D^+ D_s^{*-})=a_1(\ov B^0\to D^+
\rho^-)$, for example. However, it is worth stressing again that the above
extraction
of $f_{D_s}$ and $f_{D_s^*}$ suffers from the uncertainty of using the same
values of $a_1$ for different channels \cite{Ciuchini}.
Since the energy released to the
$DD_s$ state is smaller than that to the $D\pi$ state, $a_1$
may differ significantly in these two decay modes.

\section{Determination of \lowercase{$a_2$} and \lowercase{$a_2/a_1$} }
In principle, the magnitude of $a_2$ can be extracted directly from the decays
$B\to J/\psi K^{(*)}$ and $\ov B^0\to D^{(*)0}\pi^0(\rho^0)$ and
indirectly from the data of $B^-\to D^{(*)}\pi(\rho)$ and
$\ov B^0\to D^{(*)}\pi(\rho)$. Unfortunately, the branching ratios of the
(class-II) color-suppressed decay modes of the neutral $B$ meson are not yet
measured. Besides the form factors,
the extraction of $a_2$ from $B\to D^{(*)}\pi(\rho)$ depends on
the unknown decay constants $f_D$ and $f_{D^*}$. On the contrary, the
decay constant $f_{J/\psi}$ is well determined and the quality of the data
for $B\to J/\psi K^{(*)}$ is significantly improved over past years.
Nevertheless, the relative sign of $a_1$ and $a_2$ can be fixed by the
measured ratios $R_1,\cdots,R_4$ [cf. Eq.~(\ref{R14})] of charged to neutral
branching fractions of $B\to D^{(*)}\pi(\rho)$, and
an upper bound on $|a_2|$ can be derived from the current limit
on $\ov B^0\to D^0\pi^0$.

\subsection{Extraction of $|a_2|$ from $B\to J/\psi K^{(*)}$ }
From Eqs.~(\ref{BII}) and (\ref{term}), it is clear that $a_2$ derived
from $B\to J/\psi K$ and $B\to J/\psi K^*$ depends on the form factors
$F_1^{BK}(m^2_{J/\psi})$ and $A_{1,2}^{BK^*}(m^2_{J/\psi}),~V^{BK^*}(m^2_{J/
\psi})$. These form factors evaluated in various models are collected in
Table VI. A fit of Eq.~(\ref{BII}) to the data of ${\cal B}(B\to J/\psi K)$
(see Table VII) yields
\be \label{a2-1}
|a_2|(B\to J/\psi K)= (0.26\pm 0.02)\left( {0.70\over F_1^{BK}
(m^2_{J/\psi}) }\right).
\en
From Table VII we also see that the extracted value of $|a_2|(B\to J/\psi
K^*)$ in various models can be approximated by
\be  \label{a2-2}
|a_2|(B\to J/\psi K^*) \approx (0.21\pm 0.02)\left( {0.45\over A_1^{BK^*}
(m^2_{J/\psi}) }\right).
\en
This implies that the quantity $\sqrt{H}$ defined in Eq.~(\ref{H}) is
essentially model-independent, which can be checked explicitly.
If the factorization approximation is good, the value of $a_2$ obtained from
$J/\psi K$ and $J/\psi K^*$ states should be close to each other. This is
justified because the energy release in $B\to J/\psi K^*$ is similar to that
in $B\to J/\psi K$ and hence the nonfactorizable effects in these
two processes should be similar.
However, we learn from Table \ref{tab:a2} that only the NRSX, LF models
and the Yang's sum rule analysis meet this expectation.

In order to have a process-insensitive $a_2$, it follows from
Eqs.~(\ref{a2-1}) and (\ref{a2-2}) that the form factors $F_1^{BK}$
and $A_1^{BK^*}$ must satisfy the relation
\be
z={F_1^{BK}(m^2_{J/\psi})\over A_1^{BK^*}(m^2_{J/\psi})}\approx 1.93\,.
\en
It is evident from Table VI that the ratio $F_1^{BK}(m^2_{J/\psi})/
A_1^{BK^*}(m^2_{J/\psi})$ is close to 1.9 in the aforementioned three models.
This is also reflected in the production ratio
\be  \label{R}
R\equiv \frac{\B{(B\to J/\psi K^*)}}{\B(B\to J/\psi K)}\,.
\en
Based on the factorization approach, the predictions of $R$ in various
form-factor models are shown in Table VIII. The BSW, NS and LCSR models
in their present forms are ruled out since they predict a too large
production ratio. To get a further insight, we consider a ratio defined by
\be   \label{Z}
Z(q^2)\equiv {F_1^{BK}(q^2)\over A_1^{BK^*}(q^2)}\Bigg/
{F_1^{BK}(0)\over A_1^{BK^*}(0)},
\en
which measures the enhancement of $F_1^{BK}/ A_1^{BK^*}$ from $q^2=0$ to
finite $q^2$. $Z$ is close to unity in the BSW model and in Yang's
sum rules (see
Table VI) because $F_1^{BK}$ and $A_1^{BK^*}$ there have the same monopole
$q^2$ dependence, while in the other models $F_1^{BK}$
increases with $q^2$ faster than $A_1$. For example, the $q^2$ dependence
of $F_1^{BK^*}$ in the LF model differs from that of $A_1^{BK^*}$ by an
additional pole factor. We see from Table VI that NS, LCSR and LF models
all have similar $q^2$ behavior
\footnote{Although $F_1^{BK}$ has the same dipole $q^2$ behavior in
NRSX and LF models, its growth with $q^2$ in the former model is slightly
faster than the latter because of the smaller pole mass.}
for $Z$ with $Z(m^2_{J/\psi})\sim {\cal O}
(1.35)$. In order to accommodate the data, we need $F_1^{BK}(0)/
A_1^{BK^*}(0)\gtrsim 1.30$. However, the values of $F_1^{BK}(0)$ and
$A_1^{BK^*}(0)$
are the same in both NS and LCSR models (see Table I) and this
explains why they fail
to explain the production ratio. By contrast, although $Z\approx 1$
in the Yang's sum rules, its $F_1^{BK}(0)$ is two times as large as
$A_1^{BK^*}(0)$ so that $F_1^{BK}(m^2_{J/\psi})/
A_1^{BK^*}(m^2_{J/\psi})\approx F_1^{BK}(0)/A_1^{BK^*}(0)\approx 2$.
We thus conclude that the data of $B\to
J/\psi K^{(*)}$ together with the factorization hypothesis imply
some severe constraints on the $B\to K^{(*)}$ transition: the
form factor $F_1^{BK}$ must be larger than $A_1^{BK^*}$ by at least 30\%
at $q^2=0$ and it must grow with $q^2$ faster than the latter so
that $F_1^{BK}(m^2_{J/\psi})/A_1^{BK^*}(m^2_{J/\psi})\approx 1.9$\,.

   Since experimental studies on the the fraction of longitudinal
polarization $\Gamma_L/\Gamma$ and the parity-odd $P$--wave component or
transverse polarization $|P|^2$ measured in the
transversity basis in $B\to J/\psi K^*$ decays are available, we have
analyzed them in various models as shown in Table VIII.
At this point, it is worth emphasizing that
the generalized factorization hypothesis is a strong
assumption for the $B\to VV$ decay mode as its
general decay amplitude consists of three independent Lorentz
structures, corresponding to $S$--, $P$-- and $D$--waves or the form factors
$A_1,~V$ and $A_2$. {\it A priori},
there is no reason to expect that nonfactorizable terms weight in the
same way to $S$--, $P$-- and $D$--waves. The generalized factorization
assumption forces all the nonfactorizable terms to be the same and
channel-independent \cite{Cheng97}. Consequently, nonfactorizable effects
in the hadronic matrix elements can be lumped into the effective coefficients
$a_i$ under the generalized factorization approximation. Since the
decay $B\to J/\psi K^{(*)}$ is color suppressed and since $|c_1/c_2|\gg 1$,
it is evident from Eq.~(\ref{aeff})
that even a small amount of nonfactorized term $\chi_2$ will have a
significant impact on its decay rate. However, it is easily seen that
nonfactorizable effects are canceled out in the production ratio, the
longitudinal polarization fraction and the $P$--wave component. Therefore,
the predictions of these three quantities are the same in the generalized and
naive factorization approaches. Explicitly \cite{Gourdin,Cheng97},
\be \label{pol}
R=\,1.08\,{H\over z^2},  \qquad\quad {\Gamma_L\over\Gamma}={(a-bx)^2\over H},
\qquad \quad |P|^2={2c^2y^2\over H},
\en
where $H,a,b,c,x,y$ are defined in Eqs.~(\ref{H}) and (\ref{abc}). Numerically,
$a=3.165,\,b=1.308,\,c=0.436$. Form factors $A_2^{BK^*}$ and $V^{BK^*}$
at $q^2=m^2_{J/\psi}$ can be inferred from the
measurements of $\Gamma_L/\Gamma$ and $|P|^2$
in $B\to J/\psi K^*$. For illustration we take the central
values of the CLEO data \cite{Jessop} (see also Table VIII):
$R=1.45,~\Gamma_L/\Gamma=0.52$
and $|P|^2=0.16$. Since $z\approx 1.9$, it follows from Eq.~(\ref{pol})
that
\be
x= {A_2^{BK^*}(m_{J/\psi}^2)\over A_1^{BK^*}(m_{J/\psi}^2) }=\,1.19\,,
\qquad \quad y={V^{BK^*}(m_{J/\psi}^2)\over A_1^{BK^*}(m_{J/\psi}^2) }=\,
1.45\,.
\en
From Table VIII we see that all the model predictions for
$\Gamma_L/\Gamma$ and $|P|^2$ are in agreement with experiment
\footnote{Historically, it has been shown \cite{Gourdin} that the earlier
data of $R$ and $\Gamma_L/\Gamma$ cannot be simultaneously accounted for by
all commonly used models for form factors. In particular, all the existing
models based on factorization cannot produce a large longitudinal
polarization fraction, $\Gamma_L/\Gamma=0.74\pm 0.07$. Various possibilities
of accommodating this large $\Gamma_L/\Gamma$ via nonfactorizable effects have
been explored in \cite{Santra,Cheng96,Cheng97}.
The new CLEO \cite{Jessop} and CDF \cite{CDF}
data for $\Gamma_L/\Gamma$ are smaller than the previous values. As a
result, there exist some form-factor models which can explain all the
three quantities $R,\,\Gamma_L/\Gamma$ and $|P|^2$ (see Table VIII).}
except that the longitudinal polarization
fraction obtained in the NRSX model is slightly small. Indeed, among
the six form-factor models under consideration, the NRSX model has the
largest value of $x$ (see Table VI), $x=1.4$ which deviates most from
the value of 1.19\,, and hence the smallest
value of $\Gamma_L/\Gamma$. As noted in \cite{CT95}, some information on the
form factors $A_1^{BK^*}$ and $V^{BK^*}$ at $q^2=0$ can be inferred
from $B\to K^*\gamma$ decays.

   It is instructive to compare the predictions of the BSW and NRSX models
for $B\to J\psi K^{(*)}$ since their $B\to K^{(*)}$ form factors at $q^2=0$
are the same. Because of the dipole behavior of the form factors $F_1,V,A_2$,
the NRSX model predicts larger values for $x,y,z$ and hence smaller values
for $R,\,\Gamma_L/\Gamma$ and a larger $|P|^2$ (see Table VIII).

 In short, in order to accommodate the data of $B\to J/\psi K^{(*)}$
within the factorization framework, the form-factor models must be
constructed in such a way that
\be
&& A_2^{BK^*}(m_{J/\psi}^2)/A_1^{BK^*}(m_{J/\psi}^2)\sim 1.2\,, \non \\
&& V^{BK^*}(m_{J/\psi}^2)/A_1^{BK^*}(m_{J/\psi}^2)\sim 1.5\,, \non \\
&& F_1^{BK}(m_{J/\psi}^2)/A_1^{BK^*}(m_{J/\psi}^2)~\sim 1.9\,.
\en

  In the literature the predicted values of $F_{0,1}^{BK}(0)$ spread over a
large range. On the one hand, a large $F_{0,1}^{BK}(0)$ is preferred by
the abnormally large branching ratio of the charmless $B$ decay
$B\to \eta' K$ observed by CLEO \cite{Behrens}. On the other hand, it
cannot be too large otherwise the SU(3)-symmetry relation $F_{0,1}^{BK}(0)
=F_{0,1}^{B\pi}(0)$ will be badly broken.
There exist many model calculations of $F_{0,1}^{B\pi}
(0)$, including the lattice one, and most of them fall into the range of
0.20--0.33 (for a compilation of previous model calculations of
$F_{0,1}^{B\pi}(0)$, see e.g. \cite{Lu}).
The improved upper limit on the decay mode $\ov B^0\to
\pi^+\pi^-$, ${\cal B}(\ov B^0\to\pi^+\pi^-)<0.84\times 10^{-5}$ obtained
recently by CLEO \cite{Roy} implies $F_{0,1}^{B\pi}(0)\lesssim 0.33$ or even
smaller \cite{Cheng98}. Therefore, even after SU(3) breaking is taken into
account, it is very unlikely that $F_{0,1}^{BK}(0)$ can exceed 0.40\,.
Our best guess is that the original BSW values, $F_{0,1}^{B\pi}(0)=0.33$ and
$F_{0,1}^{B K}(0)=0.38$ \cite{BSW85,bsw} are still very plausible. Taking $F_1
^{BK}(0)=0.38$ and using the $q^2$ dependence implied by the LCSR (or
NS, LF models), we find $F_1^{BK}(m^2_{J/\psi})\approx 0.70$ and hence
$|a_2|(B\to J/\psi K)\approx 0.26\pm 0.02$ followed from Eq.~(\ref{a2-1}).

\subsection{Extraction of $a_2/a_1$ and $a_2$ from $B\to D^{(*)}\pi(\rho)$}
   The effective coefficient $a_2$ and its sign relative to $a_1$ can
be extracted from class-III decays $B^-\to D^{(*)0}\pi^-(\rho^-)$ in
conjunction with the class-I ones $\ov B^0\to D^{(*)+}\pi^-(\rho^-)$,
as the former involve
interference between external and internal $W$--emission diagrams, while
the latter proceed through the external $W$--emission. Unlike the
determination of $a_1$, there is no analogous differential semileptonic
distribution that can be related to the
color-suppressed hadronic decay via factorization. Since the decay
constants $f_D$ and $f_{D^*}$ are still unknown, the results for
$a_2/a_1$ determined from the ratios $R_{1,2}$ and $R_{3,4}$ of charged to
neutral branching fractions [see Eq.~(\ref{R14}) for the definition] are
normalized at $f_D=200$ MeV and $f_{D^*}=230$
MeV, respectively (Table IX). We see that $a_2/a_1$
varies significantly from channel to channel and its value is mainly
governed by $R_1$ and $R_3$.
\footnote{The data of $R_1,\cdots,R_4$ are taken from the Particle Data
Group (PDG) \cite{PDG}.
Recently, CLEO has reported a new measurement of $B\to D^*\pi$ and obtained
$R_3=1.55\pm 0.14\pm 0.15$ \cite{Brandenburg}, to be compared with
$R_3=1.67\pm 0.19$ employed in Table IX.}
Combining $a_2/a_1$ with Table II for $a_1$ yields the desired results
for $a_2$ as shown in Table X. It is well known that the sign of $a_2$ is
positive because of the constructive interference in $B^-\to D^{(*)0}
\pi^-(\rho^-)$, which in turn implies that the ratios $R_1,\cdots,R_4$
are greater than unity.

\subsection{Upper limit on $a_2$ from $\ov B^0\to D^0\pi^0$}
   From the last subsection we learn that the sign of $a_2/a_1$ is
fixed to be positive due to the constructive interference in the class-III
modes $B^-\to D^{(*)0}
\pi^-(\rho^-)$, but its magnitude is subject to large errors. It is thus
desirable to extract $a_2$ directly from class-II modes, e.g.
$\ov B^0\to D^{(*)0}\pi^0(\rho^0)$. Although only upper limits
on color-suppressed decays are available at present, the lowest
upper limit ${\cal B}(\ov B^0\to D^0\pi^0)<1.2
\times 10^{-4}$ \cite{Nemati} can be utilized to set a stringent
bound on $a_2$. Neglecting $W$-exchange and final-state interactions for
the moment, we obtain
\be
|a_2|(B\to D\pi)<0.29\left( {0.373\over F_0^{B\pi}(m_D^2)} \right)
\left( {200\,{\rm MeV}\over f_D}\right).
\en
The limit on $a_2$ in various form-factor models for $F_0^{B\pi}$
is shown in Table XI.

   We have argued in passing that final-state interactions (FSI) play a minor
role in hadronic $B$ decays, especially class-I modes.
In order to have a concrete estimate of FSI, we decompose the physical
amplitudes into their isospin amplitudes
\be
A(\ov B^0\to D^+\pi^-)_{\rm FSI} &=& \sqrt{2\over 3}A_{1/2}e^{i\delta_{1/2}}
+\sqrt{1\over 3}A_{3/2}e^{i\delta_{3/2}},   \non \\
A(\ov B^0\to D^0\pi^0)_{\rm FSI} &=& \sqrt{1\over 3}A_{1/2}e^{i\delta_{1/2}}
-\sqrt{2\over 3}A_{3/2}e^{i\delta_{3/2}},   \non \\
A(B^-\to D^0\pi^-)_{\rm FSI} &=& \sqrt{3}A_{3/2}e^{i\delta_{3/2}},
\en
where we have put in isospin phase shifts and assumed that inelasticity is
absent or negligible so that the isospin phase shifts are real and
the magnitude of
the isospin amplitudes $A_{1/2}$ and $A_{3/2}$ is not affected by FSI.
The isospin amplitudes are related
to the factorizable amplitudes given in Eqs.~(\ref{BI}), (\ref{BII'}) and
(\ref{BIII}) by setting $\delta_{1/2}=\delta_{3/2}=0$. Writing
\be
{\cal T} &=& {G_F\over\sqrt{2}}\,V_{cb}V_{ud}^*\,a_1(m_B^2-m_D^2)f_\pi
F_0^{BD}(m_\pi^2),   \non \\
{\cal C} &=& {G_F\over\sqrt{2}}\,V_{cb}V_{ud}^*\,a_2(m_B^2-m_\pi^2)f_D
F_0^{B\pi}(m_D^2),
\en
for color-allowed and color-suppressed tree amplitudes, respectively, it
is straightforward to show that
\be
A(\ov B^0\to D^0\pi^0)_{\rm FSI} &=& A(\ov B^0\to D^0\pi^0)+{2{\cal T}-
{\cal C}\over 3\sqrt{2}}\left( e^{i(\delta_{1/2}-\delta_{3/2})}-1\right),
\non \\
A(\ov B^0\to D^+\pi^-)_{\rm FSI} &=& A(\ov B^0\to D^+\pi^-)+{2{\cal T}-
{\cal C}\over 3}\left( e^{i(\delta_{1/2}-\delta_{3/2})}-1\right),
\en
where $A(\ov B^0\to D^0\pi^0)=-{\cal C}/\sqrt{2}$, $A(\ov B^0\to D^+\pi^-)
={\cal T}$, and we have dropped the overall phase $e^{i\delta_{3/2}}$.
Taking $a_1=1$ and $a_2=0.25$ as an illustration, we plot
in Fig.~1 the effect of FSI on $\Gamma(\ov B^0\to D\pi)$ versus
the isospin phase shift difference using the NRSX form-factor model.
We see that FSI will suppress the decay rate of
$\ov B^0\to D^+\pi^-$ slightly, but enhance that of $\ov B^0\to D^0\pi^0$
significantly, especially when $|\delta_{1/2}-\delta_{3/2}|$ is close to
the current limit $19^\circ$ [cf. Eq.~(\ref{19})]. This is understandable
because the branching
ratio of $\ov B^0\to D^0\pi^0$ in the absence of FSI is much smaller than
that of $\ov B^0\to D^+\pi^-$. Therefore, even a small amount of FSI via the
$D^+\pi^-$ intermediate state will enhance the decay rate of $\ov B^0\to D^0
\pi^0$ significantly. Fig.~2 displays the change of the upper limit of
$a_2$ in the NRSX model with respect to the phase shift
difference, where we have set $a_1=1$. Evidently, the bound on $a_2$
becomes more stringent as $|\delta_{1/2}-\delta_{3/2}|$ increases; we find
$a_2(B\to D\pi)<0.29\times (200\,{\rm MeV}/f_D)$ in the
absence of FSI and
$a_2<0.21\times (200\,{\rm MeV}/f_D)$ at $|\delta_{1/2}-\delta_{3/2}|=
19^\circ$ (see Table XI for other model predictions).

\subsection{Sign of $a_2(B\to J/\psi K^{(*)})$}
Although the magnitude of $a_2$ extracted from $B\to J/\psi K^{(*)}$ has
small errors compared to that determined from the interference effect in
$B\to D\pi(\rho)$, its sign remains unknown. Since $a_2(B\to D\pi)$ is
positive in the usual sign convention for $a_1$, it is natural to assign the
same sign to the $J/\psi K^{(*)}$ channel. It has been long advocated in
\cite{Ruckl} that the sign of $a_2(B\to J/\psi K)$ predicted by the sum rule
analysis is opposite to the above expectation. However, we
believe that a negative sign for $a_2(B\to J/\psi K)$ is very unlikely for
three main reasons: (i) Taking $|a_2(B\to J/\psi K)|=0.26$ as a representative
value and using $c_1^{\rm eff}=1.149$, $c_2^{\rm eff}=-0.325$
from Eq.~(\ref{ceff}), we
obtain two possible solutions for the nonfactorizable term $\chi_2(B\to
J/\psi K)$ [see Eq.~(\ref{aeff})]:
$\chi_2=0.18$ and $\chi_2=-0.28$. Recall that $\chi_2(B\to D\pi)$ is positive
and of order 0.15 \cite{CT98}. Though the energy release in $B\to J/\psi K$ is
somewhat smaller than that in the $D\pi$ mode, it still seems very unlikely
that $\chi_2$ will change the magnitude and in particular the sign suddenly
from the $D\pi$ channel to the $J/\psi K$ one. To make our point more
transparent, we note that $\chi_2$ has the expression:
\be
\chi_2(B\to J/\psi K)=\,\vp_8^{(BK,J/\psi)}+{a_2\over c_1}\,\vp_1^{(BK,J/
\psi)},
\en
where the parameters $\vp_8$ and $\vp_1$ are defined in Eq.~(\ref{epsilon}).
Since $c_1\gg a_2$, it is evident that $\chi_2$ is dominated by the
parameter $\vp_8$ originated from color octet-octet currents; that is, the
nonfactorized term $\chi_2$ is governed by soft gluon interactions.\footnote{
In the large-$N_c$ limit, $\vp_1$ is suppressed relative to $\vp_8$
by a factor of $N_c$ \cite{ns}. Numerically, $\vp_1(\mu)=-0.07\pm 0.03$
and $\vp_8(\mu)=0.13\pm 0.05$ at $\mu=4.6$ GeV are found in \cite{Kamal} by
extracting them from the data.
However, it has been shown in \cite{Buras} that $\vp_1(\mu)$ is not
necessarily smaller than $\vp_8(\mu)$, but this will not affect the
conclusion that $\chi_2$ is dominated by the $\vp_8$ term.}
Therefore, $|\chi_2|$ should become
smaller when the energy released to the final-state particles becomes larger,
for example, $|\chi_2(B\to D\pi)|\ll |\chi_2(D\to\ov K\pi)|$.
It is natural to expect that $|\vp_8(B\to D\pi)|\lesssim
|\vp_8(B\to J/\psi K)|\ll |\vp_8(D\to\ov K\pi)|$ and hence $|\chi_2(B\to D\pi)
|\lesssim
|\chi_2(B\to J/\psi K)|\ll |\chi_2(D\to\ov K\pi)|$ as the decay particles in
the latter channel are moving slower, allowing more time for involving
soft gluon final-state interactions. Because $\chi_2(D\to\ov K\pi)\sim
-{1\over 3}$, the solution $\chi_2(B\to J/\psi K)=0.28$ is thus not favored
by the above physical argument. (ii) Relying on a
different approach, namely, the three-scale PQCD factorization theorem,
the authors of \cite{Li} are able to explain the sign change of $\chi_2$
from $B\to D\pi$ to $D\to\ov K\pi$, though the application of PQCD to
the latter is only marginal. The same approach predicts a positive $a_2$
for $B\to J/\psi K^{(*)}$, as expected \cite{Yeh}. (iii) The existing sum rule
analysis does confirm the cancellation between the $1/N_c$ Fierz term and
$\chi_2$ for the charmed decay $D\to \ov K\pi$ \cite{Blok}, but it also
shows that the
cancellation persists even in hadronic two-body decays of $B$ mesons
\cite{BS,Ruckl,Halperin}. For example, the light-cone
QCD sum rule calculation of nonfactorizable effects in $\ov B^0\to D^0\pi^0$
in \cite{Halperin} yields a negative $\chi_2$ and $a_2$, which is in
contradiction with experiment. This means that care must be taken when
applying the sum rule analysis to the $B$ decays. Indeed, there exist
some loopholes in the conventional sum rule description of
nonleptonic two-body decays (see also the comment made in \cite{Li}), a
challenging issue we are now in progress for investigation.

\subsection{Effective $\nc$}
Since $c_1\gg c_2$, the effective coefficient $a_2$ is sensitive to the
nonfactorizable effects, and hence it is more suitable than $a_1$ for
extracting $\nc$ [strictly speaking, $(\nc)_2$]
the effective number of colors defined in Eq.~(\ref{neff}),
or the nonfactored term $\chi_2$. Although we have argued before that
$a_2(B\to J/\psi K^{(*)})\approx 0.26$ and $a_2(B\to D\pi)\lesssim a_2(
B\to J/\psi K^{(*)})$, it is safe to conclude that $a_2$ lies in the
range of 0.20--0.30. Using the renormalization scheme and scale independent
Wilson coefficients $c_1^{\rm eff}=1.149$ and $c_2^{\rm eff}=-0.325$
[cf. Eq.~(\ref{ceff})], it follows that
\be
\nc\sim (1.8-2.2),  \qquad{\rm or} \qquad \chi_2\sim (0.12-0.21),
\en
recalling that $\chi_2(D\to\ov K \pi)\sim -{1\over 3}$. Therefore,
$\nc$ for $(V-A)(V-A)$ 4-quark interactions is of order 2\,.
If $\chi_1=\chi_2$, the corresponding $a_1$ is found to be in the
range of 0.97--1.01\,.

\section{Conclusion} \label{sec:con}
Using the recent experimental data of $B\to D^{(*)}
(\pi,\rho)$, $B\to D^{(*)} D_s^{(*)}$, and $B\to J/\psi K^{(*)}$
and various model calculations on form factors, we have re-analyzed
the effective coefficients $a_1$ and $a_2$ and their ratio. Our results
are:
\begin{itemize}
\item
The extraction of $a_1$ and $a_2$ from the processes $B\to D^{(*)}D_s
^{(*)}$ and $J/\psi K^{(*)}$ is contaminated by QCD and electroweak
penguin contributions. We found that the penguin correction to the
decay amplitude is sizable for $B\to DD_s$, but only at the 4\%
level for $B\to D^*D_s,~D^{(*)}D_s^*,~J/\psi K^{(*)}$.
\item
The model-dependent extraction of $a_1$ from $B\to D^{(*)}\pi(\rho)$ is more
reliable than that from $B\to D^{(*)}D_s^{(*)}$ as the latter involve
uncertainties from penguin corrections, unknown decay constants $f_{D_s}$,
$f_{D_s^*}$ and the poor precision of the measured branching ratios.
\item
In addition to the model-dependent determination, $a_1$ has also
been extracted in a model-independent way based on the observation
that the decays $B\to D^{(*)}h$ can be related by factorization
to the measured semileptonic differential distribution of
$B\to D^{(*)}\ell\bar\nu$ at $q^2=m_h^2$. The model-independent results
$a_1(\ov B^0\to D^+\pi^-)=0.93\pm 0.10,~a_1(\ov B^0\to D^+\rho^-)=0.95\pm
0.12$ and $a_1(\ov B^0\to D^{*+}
\pi^-)=1.09\pm 0.07$ should be reliable and trustworthy.  More precise
measurements of the differential distribution are needed in order to improve
the model-independent determination of $a_1$ and to have a stringent
test of factorization.
\item
Armed with the model-independent results for $a_1$, we have extracted
heavy-to-heavy form factors from $B\to D^{(*)}\pi$: $F_{0}^{BD}(m_\pi^2)
=0.66\pm 0.06\pm 0.05$ and
$A_0^{BD^*}(m_\pi^2)=0.56\pm 0.03\pm 0.04$, where the first error is due to
the measured branching ratios and the second one due to quark-mixing matrix
elements. Form factors at other values of $q^2$ are given in Eq.~(\ref{form}).
\item
Based on the assumption that $a_1$ derived from $B\to D^{(*)}(\pi,\rho)$
and from $B\to D^{(*)}D_s^{(*)}$ is the same, it is possible to extract
the decay constants $f_{D_s}$ and $f_{D_s^*}$ in an essentially
model-independent way from the data.
We found $f_{D_s}\sim f_{D_s^*}\sim {\cal O}(260\,{\rm MeV})$ with large
errors. However, this extraction suffers from the uncertainty that we
do not know how to estimate the violation of the above assumption.
\item
By requiring that $a_2$ extracted from $J/\psi K$ and $J/\psi K^*$ channels
be similar, as implied by the factorization hypothesis,
$B\to K^{(*)}$ form factors must respect the relation
$F_1^{BK}(m^2_{J/\psi})\approx 1.9 A_1^{BK^*}(m^2_{J/\psi})$. Some existing
models in which $F_1^{BK}(0)$ is close to $A_1^{BK^*}(0)$ and/or $F_1^{BK}$
does not increase with $q^2$ faster enough than $A_1^{BK^*}$ are ruled
out. Form factors $A_2^{BK^*}$ and $V^{BK^*}$ can be inferred from the
measurements of the fraction of longitudinal polarization
and the $P$--wave component in $B\to J/\psi K^*$.
For example, the central values of the CLEO data for these two quantities
imply $A_2^{BK^*}(m_{J/\psi}^2)/A_1^{BK^*}(m_{J/\psi}^2)\approx 1.2$
and $V^{BK^*}(m_{J/\psi}^2)/A_1^{BK^*}(m_{J/\psi}^2)\approx 1.5$. We
conjecture that $F_1^{BK}(m_{J/\psi}^2)\approx 0.70$ and hence
$|a_2(B\to J/\psi K^{(*)})|\approx 0.26\pm 0.02$.
\item
We have determined the magnitude and the sign of $a_2$ from class--I and
class--III decay modes of $B\to D^{(*)}\pi(\rho)$. Unlike $a_2$ extracted
from $B\to J/\psi K$, its determination from $D^{(*)}\pi(\rho)$ channels
suffers from a further uncertainty due to the unknown decay constants $f_D$
and $f_{D^*}$.
A stringent upper limit on $a_2$ is derived from the current bound on
$\ov B^0\to D^0\pi^0$ and it is sensitive to final-state interactions.
We have argued that the sign of $a_2(B\to J/\psi K)$
should be the same as $a_2(B\to D\pi)$ and that $a_2(B\to D\pi)\lesssim
a_2(B\to J/\psi K)$.
\item
For $a_2$ in the range of 0.20--0.30, the effective number of colors
$\nc$ is in the vinicity of 2\,.

\end{itemize}

\vskip 0.5cm
\acknowledgments
This work was supported in part by the National
Science Council of R.O.C. under Grant No. NSC88-2112-M-001-006.

%%%%%%%%%%%%%%%%%%%%%%%%%%%%%%%%%%%%%%%%%%%%%%%%%%%%%%%%%%%%%%%%%%%%

\begin{table}[ht]
\caption{Form factors at zero momentum transfer and pole masses, whenever
available, in various form-factor models.
\label{tab:ff}}
\begin{center}
\begin{tabular}{lllllll}
                        &     BSW    &     NRSX      &      LF    &  NS
& Yang & LCSR    \\
\hline
 $F^{BD}_0(0)/m_{pole}$  & 0.690/6.7   & 0.58     &  0.70/7.9 & 0.636  &  &
     \\
 $F^{BD}_1(0)/m_{pole}$  & 0.690/6.264 &  0.58  &  0.70/6.59 & 0.636   &
  &    \\
 $A^{BD^*}_0(0)/m_{pole}$ &0.623/6.264 & 0.59  & 0.73/6.73 & 0.641 &
  &    \\
 $A^{BD^*}_1(0)/m_{pole}$&0.651/6.73 &  0.57  & 0.682/7.2 & 0.552  &
  &   \\
 $A^{BD^*}_2(0)/m_{pole}$&0.686/6.73 &  0.54  &0.607/7.25& 0.441 & & \\
 $V^{BD^*}(0)/m_{pole}$ &0.705/6.337& 0.76  &0.783/7.43& 0.717  &  & \\
 $F^{B\pi}_0(0)/m_{pole}$&0.333/5.73 &  0.333/5.73  &  0.26/5.7 & 0.257 &
 (see text) & 0.305  \\
 $F^{B\pi}_1(0)/m_{pole}$&0.333/5.3249& 0.333/5.3248&  0.26/5.7& 0.257 &
 0.29/5.45 & 0.305 \\
 $A^{B\rho}_0(0)/m_{pole}$&0.281/5.2789&0.281/5.2789&  0.28/5.8& 0.257 &
 (see text) & 0.372 \\
 $A^{B\rho}_1(0)/m_{pole}$&0.283/5.37&  0.283/5.37  & 0.203/5.6& 0.257 &
 0.12/5.45 & 0.261 \\
 $A^{B\rho}_2(0)/m_{pole}$&0.283/5.37&  0.283/5.37  & 0.177/6.1& 0.257 &
 0.12/6.14 & 0.223 \\
 $V^{B\rho}(0)/m_{pole}$ & 0.329/5.3249&0.329/5.3248&  0.296/--- & 0.257 &
 0.15/5.78 & 0.338 \\
 $F^{BK}_0(0)/m_{pole}$ &  0.379/5.3693&  0.379/5.3693  & 0.34/5.83& 0.295 &
 (see text) & 0.341 \\
 $F^{BK}_1(0)/m_{pole}$ &  0.379/5.41&  0.379/5.41  & 0.34/5.83& 0.295 &
 0.36/5.8 & 0.341 \\
 $A^{BK^*}_0(0)/m_{pole}$  & 0.321/5.89&  0.321/5.89  &0.32/5.83& 0.295 &
 (see text) & 0.470 \\
 $A^{BK^*}_1(0)/m_{pole}$  &  0.328/5.90 &  0.328/5.90   &0.261/5.68& 0.295
 & 0.18/6.1  & 0.337 \\
 $A^{BK^*}_2(0)/m_{pole}$  &  0.331/5.90 &  0.331/5.90   &0.235/6.11& 0.295
 &0.17/6.04 & 0.283 \\
 $V^{BK^*}(0)/m_{pole}$  & 0.369/5.41&  0.369/5.41  &0.346/10.5& 0.295 &
 0.21/5.95 & 0.458 \\
\end{tabular}
\end{center}
\end{table}
%%%%%%%%%%%%%%%%%%%%%%%%%%%%%%%%%%%%%%%%%%%%%%%%%%%%%%%%%%%%%%%%%%%%%%%%%

{\squeezetable
\begin{table}[ht]
\caption{The effective parameter $a_1$ extracted
from $\ov B^0 \to D^{(*)+} (\pi^-,\rho^-)$ using different
form-factor models. The first error comes from the experimental branching
ratios shown in the last column
and the second one from the $B$ meson lifetimes and quark-mixing
matrix elements.
\label{tab:a1Dpi} }
\begin{center}
\begin{tabular}{lcccc|c}
&    BSW     &    NRSX     &      LF   &   NS~ &
Br(\%) \cite{PDG}\\
\hline $\ov B^0\to D^+ \pi^-$
&0.89$\pm 0.06\pm 0.07$ & 1.06$\pm 0.07\pm 0.08$ &0.87$\pm 0.06\pm 0.07$
& 0.96$\pm 0.06\pm 0.07$~~~ & $0.30\pm 0.04$  \\
$\ov B^0\to D^+ \rho^-$
&0.91$\pm 0.08\pm 0.07$ & 1.06$\pm 0.09\pm 0.08$ &0.89$\pm 0.08\pm 0.07$
&0.97$\pm 0.09\pm 0.08$~~~ & $0.79\pm 0.14$ \\
$\ov B^0\to D^{*+} \pi^-$
&0.98$\pm 0.04\pm 0.08$ & 1.03$\pm 0.04\pm 0.08$ & 0.83$\pm 0.03\pm 0.06$
&0.95$\pm 0.04\pm 0.07$~~~ & $0.276\pm 0.021$ \\
$\ov B^0 \to D^{*+}\rho^-$
&0.86$\pm 0.21\pm 0.07$ & 0.92$\pm 0.23\pm 0.07$ &0.74$\pm 0.18\pm 0.06$
&0.85$\pm 0.21\pm 0.07$~~~ & $0.67\pm 0.33$ \\
\hline Average
&0.94$\pm 0.03\pm 0.07$ & 1.04$\pm 0.03\pm 0.08$ &0.85$\pm 0.03\pm 0.07$
& 0.95$\pm 0.03\pm 0.07$~~~ \\
\end{tabular}
\end{center}
\end{table} }
%%%%%%%%%%%%%%%%%%%%%%%%%%%%%%%%%%%%%%%%%%%%%%%%%%%%%%%%%%%%%%%%%%%%%%%%%%%%%%

{\squeezetable
\begin{table}[ht]
\caption{The effective parameter $a_1$ extracted
from $B \to D^{(*)}D_s^{(*)}$ using different form-factor
models. Penguin corrections to $a_1$ [see Eq.~(\ref{penguin})] are included.
The first error comes from the experimental branching ratios shown in the
last column and the second
one from the $B$ meson lifetimes and quark-mixing matrix elements. The value
of $a_1$ determined from $B\to D^{(*)}D_s$ and $B\to D^{(*)}D_s^*$ should
be multiplied by a factor of (240\,MeV/$f_{D_s}$) and (275\,MeV/$f_{D_s^*}$),
respectively.
\label{tab:a1DD}}
\begin{center}
\begin{tabular}{lcccc|c}
  &     BSW     &    NRSX     & LF  &   NS~   & Br(\%)
\cite{PDG} \\ \hline
$\ov B^0\to D^+ D_s^-$
 & $0.97\pm 0.18\pm 0.08$ & $1.12\pm 0.21\pm 0.09$ & $0.98\pm 0.18\pm 0.08$
 & $1.05\pm 0.20\pm 0.08$ ~~~ & $0.8\pm 0.3$ \\
 $B^- \to D^0 D_s^-$
 & $1.20\pm 0.18\pm 0.09$ & $1.39\pm 0.21\pm 0.11$ & $1.21\pm 0.19\pm 0.09$
 & $1.29\pm 0.20\pm 0.10$ ~~~ & $1.3\pm 0.4$ \\

 $\ov B^0\to D^{*+} D_s^-$
 & $1.29\pm 0.23\pm 0.10$ & $1.22\pm 0.22\pm 0.09$ & $1.02\pm 0.18\pm 0.08$
 & $ 1.17\pm 0.21\pm 0.09$ ~~~ & $0.96\pm 0.34$ \\
 $B^- \to D^{*0}D_s^-$
 & $1.40\pm 0.29\pm 0.11$ & $1.33\pm 0.28\pm 0.10$ & $1.01\pm 0.23\pm 0.09$
 & $1.26\pm 0.26\pm 0.10$ ~~~ & $1.2\pm 0.5$ \\

 Average & $1.17\pm 0.11\pm 0.08$ & $1.26\pm 0.11\pm 0.09$ & $1.08\pm 0.10
 \pm 0.08$ & $1.18\pm 0.11\pm 0.08$ ~~~ \\
 \hline

 $\ov B^0\to D^+ D_s^{*-}$
 & $1.29\pm 0.22\pm 0.10$ & $1.33\pm 0.33\pm 0.10$ & $1.15\pm 0.29\pm 0.09$
 & $1.29\pm 0.32\pm 0.10$ ~~~ & $1.0\pm 0.5$ \\
 $B^-\to D^0 D_s^{*-}$
 & $1.18\pm 0.26\pm 0.09$ & $1.23\pm 0.27\pm 0.10$ & $1.06\pm 0.24\pm 0.08$
 & $1.19\pm 0.26\pm 0.09$ ~~~ & $0.9\pm 0.4$ \\

 $\ov B^0\to D^{*+} D_s^{*-}$
 & $0.91\pm 0.16\pm 0.17$ & $0.98\pm 0.17\pm 0.08$ & $0.86\pm 0.15\pm 0.07$
 & $0.91\pm 0.16\pm 0.07$ ~~~ & $2.0\pm 0.7$ \\
 $B^- \to D^{*0} D_s^{*-}$
 & $1.09\pm 0.20\pm 0.08$ & $1.17\pm 0.22\pm 0.09$ & $1.03\pm 0.19\pm 0.08$
 & $1.09\pm 0.20\pm 0.08$ ~~~ & $2.7\pm 1.0$ \\

 Average & $1.05\pm 0.12\pm 0.08$ & $1.12\pm 0.12\pm 0.08$ & $0.98\pm 0.11
 \pm 0.08$ & $1.05\pm 0.11\pm 0.08$ ~~~ &  \\
\end{tabular}
\end{center}
\end{table} }

%%%%%%%%%%%%%%%%%%%%%%%%%%%%%%%%%%%%%%%%%%%%%%%%%%%%%%%%%%%%%%%%%%%%%

\begin{table}[ht]
\caption{ Values of $Y_P^{(*)}$ defined in Eq.~(\ref{Y}) in various
form-factor models.}
\begin{center}
\footnotesize
\begin{tabular}{lcccc}
  &     BSW     &    NRSX     & LF  &   NS    \\ \hline
$Y_\pi$ & 1.002 & 1.0008 & 1.0009 & 1.001  \\
$Y^*_\pi$ & 1.008 & 0.993 & 0.974 & 1.008 \\
$Y_{D_s}$ & 1.579 & 1.321 & 1.269 & 1.386 \\
$Y^*_{D_s}$ & 0.309 & 0.400 & 0.432 & 0.376 \\
\end{tabular}
\end{center}
\end{table}

%%%%%%%%%%%%%%%%%%%%%%%%%%%%%%%%%%%%%%%%%%%%%%%%%%%%%%%%%%%%%%%%%%%%%

\begin{table}[ht]
\caption{A determination of the effective parameter $a_1$
from the ratio $S_h^{(*)}$ (in units of GeV$^2$) defined in Eq.~(\ref{S}).
The data of $d{\cal B}/dq^2(\ov B^0\to D^+\ell^-\bar\nu)$ (in units of
$10^{-2}\,{\rm GeV}^{-2}$) denoted by (*) are explained in the text. The value
of $a_1$ determined from $B\to D^{(*)}D_s$ and $B\to D^{(*)}D_s^*$ should
be multiplied by a factor of (240\,MeV/$f_{D_s}$) and (275\,MeV/$f_{D_s^*}$),
respectively.
 }
\begin{center}
\begin{tabular}{l c c c | c c c}
$q^2$  & ${d\over dq^2}{\cal B}(\ov B^0\to D^+ \ell^-\bar\nu)$ & $S_h$ &
$a_1$ ~~~& ${d\over dq^2}{\cal B}(\ov  B^0\to D^{*+}\ell^-\bar\nu)$ &
$S_h^*$ & $a_1$ \\   \hline
$m_\pi^2$ & $0.35\pm 0.06$ & $0.86\pm 0.19$ & $0.93\pm 0.10$ ~ &
$0.237\pm 0.026$ & $1.16\pm 0.16$ & $1.09\pm 0.07$ \\
$m_\rho^2$ & $0.33\pm 0.06$ & $2.39\pm 0.61$ & $0.95\pm 0.12$ ~ &
$0.250\pm 0.030$ & $2.68\pm 1.36$ & $1.01\pm 0.26$ \\
$m_{D_s}^2$ & $0.29\pm 0.06^{(*)}$ & $3.24\pm 1.07$ & $1.01\pm 0.14$ ~ &
$0.483\pm 0.033$ & $2.09\pm 0.60$ & $1.22\pm 0.18$ \\
$m_{D_s^*}^2$ & $0.27\pm 0.06^{(*)}$ & $3.33\pm 1.34$ & $0.92 \pm 0.18$ ~ &
$0.507\pm 0.035$ & $4.32\pm 1.22$ & $1.05\pm 0.14$ \\
\end{tabular}
\end{center}
\end{table}
%%%%%%%%%%%%%%%%%%%%%%%%%%%%%%%%%%%%%%%%%%%%%%%%%%%%%%%%%%%%%%%%%%%%%%%%%%%%%%
\begin{table}[ht]
\caption{Form factors $F_1^{BK}$, $A_1^{BK^*},A_2^{BK^*}$, $V^{BK^*}$ and the
ratio $Z$ [see Eq.~(\ref{Z})] at
$q^2=m^2_{J/\psi}$ in various form-factor models. }
\begin{center}
\begin{tabular}{lllllll}
                        &     BSW    &     NRSX      &      LF    &  NS
& Yang & LCSR    \\
\hline
 $F^{BK}_1(m^2_{J/\psi})$ & 0.56 & 0.84 & 0.66 & 0.52 & 0.50 & 0.62 \\
 $A^{BK^*}_1(m^2_{J/\psi})$  & 0.45 & 0.45 & 0.37 & 0.39 & 0.24 & 0.43 \\
 $A^{BK^*}_2(m^2_{J/\psi})$  & 0.46 & 0.63 & 0.43 & 0.48 & 0.31 & 0.45 \\
 $V^{BK^*}(m^2_{J/\psi})$ & 0.55 & 0.82 & 0.42 & 0.51 & 0.40 & 0.86 \\
 $Z(m_{J/\psi}^2)$ & 1.08 & 1.60 & 1.36 & 1.32 & 1.04 & 1.40 \\
\end{tabular}
\end{center}
\end{table}

%%%%%%%%%%%%%%%%%%%%%%%%%%%%%%%%%%%%%%%%%%%%%%%%%%%%%%%%%%%%%%%%%%%%%%%%%

{\squeezetable
\begin{table}[t] \caption{The effective parameter $|a_2|$ extracted
from $B\to J/\psi K^{(*)}$ using different form-factor models.
Experimental branching ratios are taken from the Particle Data Group.
\label{tab:a2}}
\begin{center}
\footnotesize
\begin{tabular}{lcccccc|c}
&   BSW    &  NRSX     &      LF       &    NS &Yang   & LCSR~~ &
Br($10^{-3}$)\cite{PDG} \\
\hline $B^+\to J/\psi K^+$
&0.34$\pm$0.03&0.23$\pm$0.02&0.29$\pm$0.03 &
 0.37$\pm$0.03&0.38$\pm$0.03&0.30$\pm$0.03 ~~ & $0.99\pm 0.10$  \\
$B^0\to J/\psi K^0$
&0.33$\pm$0.03&0.22$\pm$0.02&0.28$\pm$0.03 &
 0.36$\pm$0.04&0.37$\pm$0.04&0.30$\pm$0.03 ~~ & $0.89\pm 0.12$ \\
        Average
& 0.33$\pm$0.03&0.22$\pm$0.02&0.29$\pm$0.02&
  0.36$\pm$0.03&0.37$\pm$0.03&0.30$\pm$0.03 ~~  \\
\hline
$B^+\to J/\psi K^{*+}$
&0.20$\pm$0.02&0.22$\pm$0.03&0.26$\pm$0.03&
 0.25$\pm$0.03&0.40$\pm$0.05&0.20$\pm$0.02 ~~ & $1.47\pm 0.27$ \\
$B^0\to J/\psi K^{*0}$
&0.20$\pm$0.02&0.22$\pm$0.02&0.26$\pm$0.03&
 0.25$\pm$0.03&0.40$\pm$0.04&0.20$\pm$0.02 ~~ & $1.35\pm 0.18$ \\
        Average
&0.20$\pm$0.02&0.22$\pm$0.02&0.26$\pm$0.02&
 0.25$\pm$0.02&0.40$\pm$0.04&0.20$\pm$0.02 ~~ \\
\end{tabular}
\end{center}
\end{table} }
%%%%%%%%%%%%%%%%%%%%%%%%%%%%%%%%%%%%%%%%%%%%%%%%%%%%%%%%%%%%%%%%%%%%%%%%%
\begin{table}[ht]
\caption{The ratio of pseudoscalar to vector meson production $R$, the
longitudinal polarization fraction $\Gamma_L/\Gamma$, and the $P$--wave
component $|P|^2$ in $B\to J/\psi K^{(*)}$ decays calculated
in various form-factor models using the factorization hypothesis.
 \label{tab:rgp}}
\begin{center}
\footnotesize
\begin{tabular}{lcccccccc}
& & & & & & & \multicolumn{2}{c}{Experiment}  \\\cline{8-9}
& \raisebox{1.5ex}[0cm][0cm]{BSW} & \raisebox{1.5ex}[0cm][0cm]{NRSX} &
\raisebox{1.5ex}[0cm][0cm]{LF} & \raisebox{1.5ex}[0cm][0cm]{NS} &
\raisebox{1.5ex}[0cm][0cm]{Yang} & \raisebox{1.5ex}[0cm][0cm]{LCSR} &
CLEO \cite{Jessop} & CDF \cite{CDF}  \\
\hline
  $R$        & 4.15 & 1.58  & 1.79 & 3.15 & 1.30 & 3.40 & $1.45\pm 0.26$
& $1.53\pm 0.32$ \\
$\Gamma_L/\Gamma$& 0.57 & 0.36  & 0.53 & 0.48 & 0.42 & 0.47 &
$0.52\pm 0.08$ & $0.65\pm 0.11$ \\
 $|P|^2$         & 0.09 & 0.24  & 0.09 & 0.12 & 0.19 & 0.23 &
$0.16\pm 0.09$ & --- \\
\end{tabular}
\end{center}
\end{table}
%%%%%%%%%%%%%%%%%%%%%%%%%%%%%%%%%%%%%%%%%%%%%%%%%%%%%%%%%%%%%%%%%%%%%%%%%%%%%%%%%%%
\begin{table}[ht]
\caption{Extraction of $a_2/a_1$ from $B\to D^{(*)}\pi(\rho)$ decays
in various form-factor models. The values of $a_2/a_1$
determined from the ratios $R_{1,2}$ and $R_{3,4}$ of charged to neutral
branching fractions [see Eq.~(\ref{R14}) for the definition]
should be multiplied by a
factor of $(200\,{\rm MeV}/f_D)$ and $(230\,{\rm MeV}/f_{D^*})$,
respectively.
\label{tab:a2a1pdg}}
\begin{center}
\footnotesize
\begin{tabular}{ccccccc|c}
       &     BSW     &    NRSX    &      LF    &      NS     & Yang &
 LCSR  & Expt. \cite{PDG} \\ \hline
 $R_1$ & 0.30$\pm$0.11 &0.26$\pm$0.10&0.40$\pm$0.15&0.39$\pm$0.15&
 0.36$\pm$0.16& 0.33$\pm$0.12~~ & $1.77\pm 0.29$  \\
 $R_2$ & 0.61$\pm$0.33&0.46$\pm$0.25&0.58$\pm$0.31&0.52$\pm$0.32&
 1.07$\pm$0.58& 0.41$\pm$0.22~~ & $1.69\pm 0.38$ \\
 Average & 0.34$\pm$0.11 & 0.28$\pm$0.09 & 0.43$\pm$0.13 & 0.43$\pm$0.13
 & 0.40$\pm$0.13 & 0.35$\pm$0.11~~ \\
\hline
 $R_3$ & 0.23$\pm$0.07& 0.19$\pm$0.06&0.31$\pm$0.09&0.28$\pm$0.08&
 0.27$\pm$0.08& 0.24$\pm$0.07~~ & $1.67\pm 0.19$  \\
 $R_4$ & 0.55$\pm$0.45&0.64$\pm$0.52&0.85$\pm$0.70&0.74$\pm$0.61&
 1.47$\pm$1.20& 0.61$\pm$0.50~~ & $2.31\pm 1.23$  \\
 Average&0.24$\pm$0.07&0.19$\pm$0.06&0.32$\pm$0.09&0.29$\pm$0.08&
 0.28$\pm$0.08& 0.25$\pm$0.07~~ & \\
\end{tabular}
\end{center}
\end{table}
%%%%%%%%%%%%%%%%%%%%%%%%%%%%%%%%%%%%%%%%%%%%%%%%%%%%%%%%%%%%%%%%%%%%%%%%%%%%%%%%%

\begin{table}[ht]
\caption{The effective coefficient $a_2$ extracted from the analyses of
$R_i$ and $\ov B^0\to D^{(*)+} \pi^-(\rho^-)$. The
values determined from $R_{1,2}$ and $R_{3,4}$ should be multiplied by a
factor of $(200\,{\rm MeV}/f_D)$ and $(230\,{\rm MeV}/f_{D^*})$,
respectively.}
\begin{center}
\footnotesize
\begin{tabular}{lcccccc}
 &    BSW    &    NRSX   &      LF     &      NS     & Yang        &LCSR \\
 \hline
 $R_1\ \&\ \ov B^0\to D^+ \pi^-$
 &0.27$\pm$0.10&0.27$\pm$0.10&0.35$\pm$0.13&0.38$\pm$0.14&0.35$\pm$0.13&0.32$
 \pm$0.12\\
 $R_2\ \&\ \ov B^0\to D^+ \rho^-$
 &0.55$\pm$0.31&0.49$\pm$0.27&0.51$\pm$0.28&0.58$\pm$0.32&1.04$\pm$0.57&0.40$
 \pm$0.22\\
 Average & 0.30$\pm$0.10 & 0.30$\pm$0.10 & 0.38$\pm$0.12 & 0.41$\pm$0.13
 & 0.38$\pm$0.13 & 0.33$\pm$0.11  \\
 \hline
 $R_3\ \&\ \ov B^0\to D^{*+} \pi^-$
 &0.22$\pm$0.07&0.19$\pm$0.06&0.26$\pm$0.08&0.27$\pm$0.08&0.26$\pm$0.08&0.23$
 \pm$0.07\\
 $R_4\ \&\ \ov B^0 \to D^{*+}\rho^-$
 &0.47$\pm$0.41&0.59$\pm$0.50&0.63$\pm$0.54&0.63$\pm$0.54&1.24$\pm$1.07&0.52$
 \pm$0.44\\
 Average
 &0.23$\pm$0.07&0.20$\pm$0.06&0.26$\pm$0.08&0.28$\pm$0.08&0.26$\pm$0.08&0.24$
 \pm$0.07\\
\end{tabular}
\end{center}
\end{table}

%%%%%%%%%%%%%%%%%%%%%%%%%%%%%%%%%%%%%%%%%%%%%%%%%%%%%%%%%%%%%%%%%%%%

\begin{table}[ht]
\caption{The upper limit on the effective coefficient $a_2$ [multiplied by
(200 MeV/$f_D$)] inferred from the decay $\ov B^0\to D^0\pi^0$ in the
absence and presence of final-state interactions characterized by the
isospin phase shift difference $\Delta=|\delta_{1/2}-\delta_{3/2}
|_{B\to D\pi}$. }
\begin{center}
\footnotesize
\begin{tabular}{lcccccc}
 & BSW    &    NRSX   &      LF     &      NS     & Yang        &LCSR \\
 \hline
$a_2$ (with $\Delta=0^\circ$) & 0.29 & 0.29 & 0.38 & 0.41 & 0.38 & 0.34 \\
$a_2$ (with $\Delta =19^\circ$) & 0.17 & 0.21 & 0.21 & 0.26 & 0.24 & 0.22 \\
\end{tabular}
\end{center}
\end{table}

%%%%%%%%%%%%%%%%%%%%%%%%%%%%%%%%%%%%%%%%%%%%%%%%%%%%%%%%%%%%%%%%%%%%

\begin{figure}[ht]
\vspace{1cm}
\hskip 3.8cm
  \psfig{figure=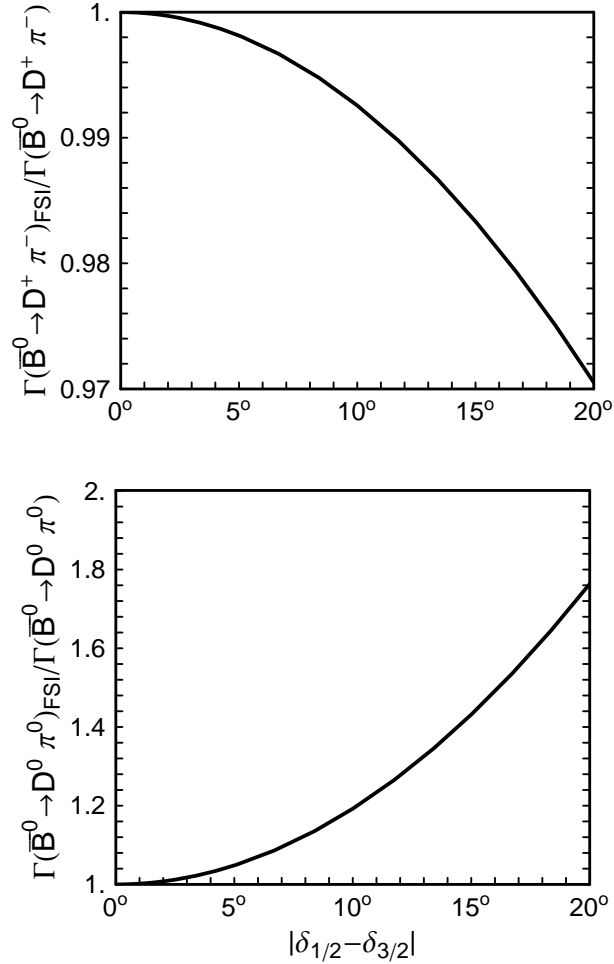,width=8cm}
\vspace{1.0cm}
    \caption[]{\small The ratio of $\Gamma(\ov B^0\to D\pi)$ in the presence
of final-state interactions (FSI) to that without FSI versus
the isospin phase-shift difference. The
calculation is done in the NRSX model \cite{NRSX}.}
\end{figure}

\begin{figure}[ht]
\vspace{1cm}
\hskip 3.8cm
  \psfig{figure=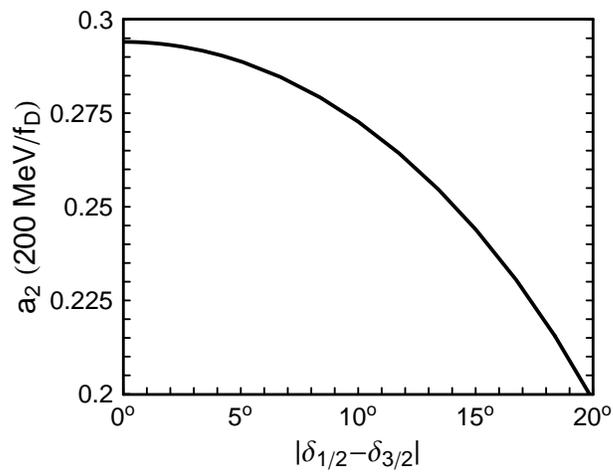,width=8cm}
\vspace{1.0cm}
    \caption[]{\small The upper bound of the effective coefficient $a_2$
multiplied by $(200\,{\rm MeV}/f_D)$
derived from the current limit on $\ov B^0\to D^0\pi^0$ using the NRSX
model \cite{NRSX} versus the isospin phase-shift difference. }
\end{figure}

\end{document}